\documentclass[11pt,twoside,a4paper]{article}
\makeatletter\if@twocolumn\PassOptionsToPackage{switch}{lineno}\else\fi\makeatother

      \makeatletter
\usepackage{wrapfig}
\newcounter{aubio}

\long\def\bioItem{%
\@ifnextchar[{\@bioItem}{\@@bioItem}}

\long\def\@bioItem[#1]#2#3{
 \stepcounter{aubio}
 \expandafter\gdef\csname authorImage\theaubio\endcsname{#1}
 \expandafter\gdef\csname authorName\theaubio\endcsname{#2}
 \expandafter\gdef\csname authorDetails\theaubio\endcsname{#3}
}

\long\def\@@bioItem#1#2{
 \stepcounter{aubio}
 \expandafter\gdef\csname authorName\theaubio\endcsname{#1}
 \expandafter\gdef\csname authorDetails\theaubio\endcsname{#2}
}

\newcommand{\checkheight}[1]{%
  \par \penalty-100\begingroup%
  \setbox8=\hbox{#1}%
  \setlength{\dimen@}{\ht8}%
  \dimen@ii\pagegoal \advance\dimen@ii-\pagetotal
  \ifdim \dimen@>\dimen@ii
    \break
  \fi\endgroup}

\def\printBio{%
  \@tempcnta=0
   \loop
     \advance \@tempcnta by 1
     \def\aubioCnt{\the\@tempcnta}
     \setlength{\intextsep}{0pt}%
     \setlength{\columnsep}{10pt}%
     \expandafter\ifx\csname authorImage\aubioCnt\endcsname\relax%
      \else%
       \checkheight{\includegraphics[height=1.25in,width=1in,keepaspectratio]{\csname authorImage\aubioCnt\endcsname}}
        \begin{wrapfigure}{l}{25mm}
         \includegraphics[height=1.25in,width=1in,keepaspectratio]{\csname authorImage\aubioCnt\endcsname}
        \end{wrapfigure}\par
      \fi
     \noindent\textbf{\csname authorName\aubioCnt\endcsname}\csname authorDetails\aubioCnt\endcsname \par\bigskip
      \ifnum\@tempcnta < \theaubio
   \repeat
   }
\makeatother

\usepackage{amsfonts,amssymb,amsbsy,latexsym,amsmath,tabulary,graphicx,times,xcolor}

\usepackage[T1]{fontenc}

\usepackage{caption,fancyhdr,abstract,textcase}
\usepackage[paperheight=297mm,paperwidth=210mm,margin=30mm,left=30mm,right=30mm,top=30mm]{geometry}
\usepackage{textcase}
\makeatletter
\def\oupIndent{1pt}
\def\author#1{\gdef\@author{\hskip-\dimexpr(\tabcolsep)\hskip1pt\parbox{\dimexpr\textwidth-1pt}{\centering{\fontsize{13pt}{15.6pt}\selectfont  #1}}}}

\def\title#1{\gdef\@title{\vspace*{-3pc}\bfseries\centering\ifx\@articleType\@empty\else\@articleType\\\fi {\fontfamily{ppl}\fontsize{20pt}{24pt}\selectfont\MakeTextUppercase{ #1} \vspace*{24pt}}}}

\fancypagestyle{headings}{\fancyhf{}\fancyfoot[R]{}}

\fancypagestyle{plain}{\fancyhf{}\fancyfoot[R]{}}

\let\@articleType\@empty \def\articletype#1{\gdef\@articleType{{\normalfont\underline{#1}}}}

\def\abstractname{\textbf{\textit{A\small{BSTRACT}}}}

\renewenvironment{abstract} {\trivlist\item[]\leftskip\oupIndent\par\vskip4pt\noindent{\fontsize{13pt}{15.6pt}\selectfont\textit{\scshape\abstractname}}\mbox{\null}\vspace{5pt}\\ \itshape\fontsize{10pt}{12pt}\selectfont}{\noindent\endtrivlist}

\usepackage[explicit]{titlesec}
\def\NormalBaseline{\def\baselinestretch{1.1}}
\titleformat{\section}[hang]{\NormalBaseline\filright\large\scshape\bfseries\fontsize{14}{16.8}\selectfont}
{\fontsize{14}{16.8}\selectfont\thesection.}
{2pt}
{#1}
[]
\titleformat{\subsection}[hang]{\NormalBaseline\filright\bfseries\fontsize{12}{14.4}\selectfont}
{\thesubsection.}
{2pt}
{#1}
[]
\titleformat{\subsubsection}[hang]{\NormalBaseline\filright\bfseries\fontsize{11}{13.2}\selectfont}
{\thesubsubsection.}
{2pt}
{#1}
[]
\titleformat{\paragraph}[runin]{\NormalBaseline\filright\itshape\fontsize{11}{13.2}\selectfont}
{\theparagraph}
{2pt}
{#1}
[\unskip.]
\titleformat{\subparagraph}[runin]{\NormalBaseline\filright\fontsize{11}{13.2}\selectfont}
{\thesubparagraph}
{2pt}
{#1}
[\unskip.]

\titlespacing{\section}{0pt}{1.5\baselineskip}{.2\baselineskip}
\titlespacing{\subsection}{0pt}{1.5\baselineskip}{.2\baselineskip}
\titlespacing{\subsubsection}{0pt}{1.5\baselineskip}{.2\baselineskip}
\titlespacing{\paragraph}{0pt}{.5\baselineskip}{10pt}
\titlespacing{\subparagraph}{0pt}{.5\baselineskip}{10pt}

\makeatother

\date{}

\usepackage{url,multirow,morefloats,floatflt,cancel,tfrupee}
\makeatletter

\AtBeginDocument{\@ifpackageloaded{textcomp}{}{\usepackage{textcomp}}}
\makeatother
\usepackage{colortbl}
\usepackage{xcolor}
\usepackage{pifont}
\usepackage[nointegrals]{wasysym}
\makeatletter

\def\mcWidth#1{\csname TY@F#1\endcsname+\tabcolsep}

\def\cAlignHack{\rightskip\@flushglue\leftskip\@flushglue\parindent\z@\parfillskip\z@skip}
\def\rAlignHack{\rightskip\z@skip\leftskip\@flushglue \parindent\z@\parfillskip\z@skip}

\usepackage{ifxetex}
\ifxetex\else\if@twocolumn\usepackage{dblfloatfix}\fi\fi

\AtBeginDocument{
\expandafter\ifx\csname eqalign\endcsname\relax
\def\eqalign#1{\null\vcenter{\def\\{\cr}\openup\jot\m@th
  \ialign{\strut$\displaystyle{##}$\hfil&$\displaystyle{{}##}$\hfil
      \crcr#1\crcr}}\,}
\fi
}

\AtBeginDocument{%
  \@ifpackageloaded{endfloat}%
   {\renewcommand\efloat@iwrite[1]{\immediate\expandafter\protected@write\csname efloat@post#1\endcsname{}}}{\newif\ifefloat@tables}%
}%

\def\BreakURLText#1{\@tfor\brk@tempa:=#1\do{\brk@tempa\hskip0pt}}
\let\lt=<
\let\gt=>
\def\processVert{\ifmmode|\else\textbar\fi}

\@ifundefined{subparagraph}{
\def\subparagraph{\@startsection{paragraph}{5}{2\parindent}{0ex plus 0.1ex minus 0.1ex}%
{0ex}{\normalfont\small\itshape}}%
}{}

\newcommand\role[1]{\unskip}
\newcommand\aucollab[1]{\unskip}

\@ifundefined{tsGraphicsScaleX}{\gdef\tsGraphicsScaleX{1}}{}
\@ifundefined{tsGraphicsScaleY}{\gdef\tsGraphicsScaleY{.9}}{}
\def\checkGraphicsWidth{\ifdim\Gin@nat@width>\linewidth
	\tsGraphicsScaleX\linewidth\else\Gin@nat@width\fi}

\def\checkGraphicsHeight{\ifdim\Gin@nat@height>.9\textheight
	\tsGraphicsScaleY\textheight\else\Gin@nat@height\fi}

\def\fixFloatSize#1{}
\let\ts@includegraphics\includegraphics

\def\inlinegraphic[#1]#2{{\edef\@tempa{#1}\edef\baseline@shift{\ifx\@tempa\@empty0\else#1\fi}\edef\tempZ{\the\numexpr(\numexpr(\baseline@shift*\f@size/100))}\protect\raisebox{\tempZ pt}{\ts@includegraphics{#2}}}}

\AtBeginDocument{\def\includegraphics{\@ifnextchar[{\ts@includegraphics}{\ts@includegraphics[width=\checkGraphicsWidth,height=\checkGraphicsHeight,keepaspectratio]}}}

\DeclareMathAlphabet{\mathpzc}{OT1}{pzc}{m}{it}

\def\URL#1#2{\@ifundefined{href}{#2}{\href{#1}{#2}}}

\def\UrlOrds{\do\*\do\-\do\~\do\'\do\"\do\-}%
\g@addto@macro{\UrlBreaks}{\UrlOrds}

\@ifundefined{quoteAttrib}
	{}
	{}

\@ifundefined{titlequoteAttrib}
	{}{}

\newenvironment{title-quote}
	{\list{}{\fontsize{10pt}{12pt}\selectfont\leftmargin.5in\itshape\rightmargin\leftmargin}%
  \item\relax}
  {\endlist}

\makeatother

\usepackage[numbers,sort&compress]{natbib}

\usepackage{supertabular}
\usepackage{graphicx}
\usepackage{longtable}
\usepackage{amsmath}
\usepackage{tabularx}

\usepackage{caption}
\usepackage{subcaption}

\usepackage{chronology}
\usepackage{algorithmic}
\usepackage[]{algorithm2e}
\usepackage{amssymb}
\usepackage{float}
\usepackage{threeparttable}

\usepackage{array}
\usepackage{subcaption}
\usepackage{multirow}
\usepackage{rotating}
\usepackage{pdflscape}

\usepackage{amssymb}

\begin{document}

\title{\huge{QoS-Driven Job Scheduling: Multi-Tier Dependency Considerations}}
\author{\Large{Husam Suleiman~and~Otman Basir}\\
\large{Department of Electrical and Computer Engineering, University of Waterloo}\\
}

\def\journalTitle{International Journal on Cloud Computing: Services and Architecture (IJCCSA)}

\maketitle

\begin{abstract}
For a cloud service provider, delivering optimal system performance while fulfilling Quality of Service (QoS) obligations is critical for maintaining a viably profitable business.
This goal is often hard to attain given the irregular nature of cloud computing jobs. These jobs expect high QoS on an on-demand fashion, that is on random arrival.
To optimize the response to such client demands, cloud service providers organize the cloud computing environment as a multi-tier architecture. Each tier executes its designated tasks and passes the job to the next tier; in a fashion similar, but not identical, to the traditional job-shop environments.
An optimization process must take place to schedule the appropriate tasks of the job on the resources of the tier, so as to meet the QoS expectations of the job.
Existing approaches employ scheduling strategies that consider the performance optimization at the individual resource level and produce optimal single-tier driven schedules. Due to the sequential nature of the multi-tier environment, the impact of such schedules on the performance of other resources and tiers tend to be ignored, resulting in a less than optimal performance when measured at the multi-tier level.

In this paper, we propose a multi-tier-oriented job scheduling and allocation technique. The scheduling and allocation process is formulated as a problem of assigning jobs to the resource queues of the cloud computing environment, where each resource of the environment employs a queue to hold the jobs assigned to it.
The scheduling problem is NP-hard, as such a biologically inspired genetic algorithm is proposed. The computing resources across all tiers of the environment are virtualized in one resource by means of a single queue virtualization. A chromosome that mimics the sequencing and allocation of the tasks in the proposed virtual queue is proposed. System performance is optimized at this chromosome level. Chromosome manipulation rules are enforced to ensure task dependencies are met. The paper reports experimental results to demonstrate the performance of the proposed technique under various conditions and in comparison with other commonly used techniques.
\end{abstract}
\emph{\textbf{KEYWORDS}}\\
\emph{\small{Cloud Computing, Task Scheduling and Allocation, QoS Optimization, Load Balancing, Genetic Algorithms}}

\section{Introduction}
\label{sec:introduct}

The advent of cloud computing has emerged as one of the latest revolutions of computing paradigms \cite{NIST_CC, An_analysis_LB2016, cloudBigPic_2015, Vinay2016}. It leverages a set of existing technologies and computing resources pooled in a cloud data center. Clients utilize cloud resources to perform complex tasks that are not easily achievable by their own infrastructure. Such resources are broadly accessed and provided as a service to clients on-demand, thus mitigate the complexity and time associated with the purchase and deployment of a traditional physical infrastructure at the client's side.

Typically, cloud computing environments experience variant workloads that entail client jobs of different QoS expectations, tardiness allowances, and computational demands. Jobs can be delay-sensitive and tightly coupled with client satisfactions, and thus cannot afford SLA violation costs. Such workload variations often occur within a short period of time and are not easily predictable, causing system bottlenecks and thus execution difficulties on cloud resources to fulfill such expectations~\cite{nextGenerCl2018}. It is imperative that a cloud service provider efficiently accommodates and responds to such demands in a timely manner, so that client experience and system performance are optimized.

Thus, the scheduling in cloud computing has become a driving theme to support a scalable infrastructure that formulates optimal workload schedules on cloud resources and mitigates potential SLA violation penalties~\cite{Mustafa2015, AbdelzahirMap2015}. The conundrum of a cloud service provider resolves around conciliating these conflicting objectives. A service provider may adopt admission control mechanisms to drop extra incoming jobs, however the likelihood of SLA violations and thus dissatisfied clients increase, which thus incurs SLA penalties on the client and service provider. In contrast, a service provider may often over-allocate resources to distinctly meet the incremental client demands and thus alleviate SLA violations, however it runs the risk of increasing the operational cost and leaving resources under-utilized.

A major limitation in schedulers of existing approaches is that they often optimize the performance of schedules at the individual resource level of a single-tier environment. However, it is typical that formulating schedules in a complex multi-tier cloud environment is harder than a traditional single-tier environment because of dependencies between the tiers. A performance degradation in a tier would propagate to negatively affect the performance of schedules in subsequent (dependent) tiers, thus causing the SLA violation penalties and likelihood of dissatisfied clients to increase.

Overall, such schedulers in their optimization strategies fail to capture QoS expectations and their associated penalties in a multi-tier environment. This paper presents a penalty-based multi-tier-driven load management approach that contemplates the impact of schedules in a tier on the performance of schedules constructed in subsequent tiers, thus optimizes the performance globally at the multi-tier level of the environment. The proposed approach accounts for tier dependencies to mitigate the potential of shifting and escalation of SLA violation penalties when jobs progress through subsequent tiers. Because the scheduling problem is NP-hard, a biologically inspired genetic algorithm supported with virtualized and segmented queue abstractions are proposed to efficiently seek (near-)optimal schedules at the multi-tier level, in a reasonable time.

\section{Background and Related Work}
\label{sec:backRelated}

Scheduling and allocation of jobs have been presented in the literature among the challenging problems in cloud computing for the past few years~\cite{Nuaimi2012, TaxonThakur2017, A_Survey_on_Scheduling_2014, heurBalanc2015}. Jobs are to be effectively scheduled and consolidated on fewer resources to deliver better system performance. Existing approaches investigate the problem from various perspectives, mostly tackled in a single-tier environment subject to common conflicting optimization objectives. The makespan and response time of jobs, as well as the resource utilization are typically the performance optimization metrics used to assess the efficacy of service delivery in achieving better user experience/satisfaction and SLA guarantees. Because the scheduling problem is NP-hard, the efficacy of scheduling approaches depends not only on fulfilling client demands and QoS obligations, but also on optimizing system performance.

Existing approaches employ different tardiness cost functions to quantify SLA violation penalties, so as to optimize the performance of schedules and mitigate their associated penalties. Chi \emph{et al}.~\cite{SLAtree1} and Moon \emph{et al}.~\cite{SoftHardSLA} adopt a stepwise function to represent different levels of SLA penalties. However, the stepwise function does not exactly reflect QoS penalty models required to tackle SLA violations of real systems. This function would typically incur a sudden change in the SLA penalty (increment/decrement from a level to another) when a slight variation in the job's completion time occurs at the transient-edge of two consecutive steps of the function, which is inaccurate. In addition, a fixed penalty level would be constantly held for each period of SLA violations, which thus inaccurately incurs equal SLA penalties for different service violation times in the same step-period. Also, formulating the cost value of each penalty level with respect to SLA violation times is still to be precisely tackled.

In addition, Stavrinides \emph{et al}.~\cite{SaasWork2017} use a linear monetary cost function to quantify multiple penalty layers (categories) of SLA violations. The tardiness metric, represented by the completion time of client jobs, is employed to calculate the cost incurred from the different layers of SLA violations. They investigate the effect of workloads of different computational demands on the performance of schedules in a single-tier environment, focusing on fair billing and meeting QoS expectations of clients. However, the linear function would not reflect the monetary cost of SLA violations in real systems, thus the performance and optimality of schedules formulated based on such cost calculations would be affected.

Furthermore, improved Min-Min and Max-Min scheduling are widely employed to tackle the problem by producing schedules at the individual resource level of the tier. Rajput \emph{et al}.~\cite{GA_MinMin_2016} and Chen \emph{et al}.~\cite{MinMin_Priority_2013} present Min-Min based scheduling algorithms to minimize the makespan of jobs and increase the resource utilization in a single-tier environment. Generally, a Min-Min approach schedules the job with the minimum completion time on the resource that executes the job at the earliest opportunity, yet negatively affects the execution of jobs with larger completion times~\cite{MinMin_2015}. In contrast, a Max-Min based approach typically utilizes powerful resources to speedup the execution of jobs with the maximum completion times, however produces poor average makespan~\cite{MaxMin_2014}.

In their optimization strategies, the Min-Min and Max-Min based approaches rely primarily on the computational demands of jobs to produce optimal schedules at the resource level. They fail to produce minimum penalty schedules that accurately account for QoS obligations of jobs at the multi-tier level, which would negatively impact provider's SLA commitments. In addition, such approaches do not consider tier dependencies of a multi-tier cloud environment, thus SLA violation penalties of schedules at the resource level would propagate to escalate in subsequent tiers, which would negatively impact system performance.

Some approaches focus on balancing the workloads among resources, as well as employing different strategies to speedup job executions~\cite{Bianca_2004, Mor_98}. Maguluri \emph{et al}.~\cite{unknownMaguluri2014} present a throughput-optimal algorithm that tackles the execution of jobs with unknown sizes. However, a throughput-based scheduling generally disregards the actual job running times in resources, and instead, focuses on queue lengths measured by the number of jobs, which is not necessarily accurate.

Redundancy-based strategies are also adopted and proven to speedup the execution of jobs~\cite{ShedRedundGARDNER2017, Mor2017ReplicRedundant}. For instance, Nahir \emph{et al}.~\cite{replica_LB2016} present a replication-based balancing algorithm that aims at minimizing the queueing overhead and the job's response time. Multiple copies (replicas) of each client's job are created and distributed on resource queues of a tier. Once a copy of the job completes the execution from a resource, other copies are deleted from the other resource queues. In addition, Kristenet \emph{et al.}~\cite{Mor2016RedundantD1, Mor2016RedundantD2} present the power of \emph{d} choices for redundancy to send copies of a job to only \emph{d} resources selected at random, so as to reduce the number of duplicated jobs in resource queues of the tier.

However, the optimization strategy of replication-based approaches does not employ the different QoS obligations and demands of jobs, thus, would not mitigate SLA violation penalties. If the mechanisms of admission control and resource over-allocation are not adopted, a replication-based approach might overload resource queues of tiers with a significant amount of jobs. Thus, the scheduler would potentially experience difficulties in managing the execution of such workloads to meet such QoS obligations at the multi-tier level.

Similar balancing approaches are widely adopted such as Least Connection (LC) weighted algorithms, Round Robin (RR) weighted algorithms~\cite{WRR_2014}, Random selection, and Shortest-Queue~\cite{JSQ_delay_2017, Randomized_JSQ_2016}. These balancing approaches are provided as a service by popular cloud providers such as Windows Azure, Amazon ELB, and HP-CLB~\cite{LBaaS2015}. Also, Wang \emph{et al}.~\cite{JIQ_2018} and Lu \emph{et al}.~\cite{JIQ_multiDispatcher_2017} present the Join-Idle-Queue (JIQ) balancing algorithm that assigns incoming jobs to only idle resource queues in a single-tier environment. Multiple dispatchers are employed to hold incoming jobs; each dispatcher keeps IDs of idle resources in the tier.

However, the JIQ-based balancing algorithm does not account for QoS expectations of jobs when a scheduling decision is made. Thus, high priority and delay-intolerant jobs might have to wait in a dispatcher to get an idle resource, while simultaneously some other delay-tolerant jobs in another dispatcher have already got idle resources for execution. In a complex multi-tier environment, the former balancing approaches would produce schedules that are poor in performance because they neither effectively reflect the system state nor account for dependencies between the tiers, and thus would not accurately meet the different QoS obligations of clients.

Furthermore, resource over-allocation is a viable option proven to provide high system performance, meet client demands, and mitigate SLA violations. Typically, clients negotiate with the service provider to submit estimates on the execution/completion times of their jobs. However, such estimates often tend to be either underestimated or inaccurate. For this purpose, Reig \emph{et al}.~\cite{predict2010} present an analytical predictor to infer job information and accordingly decide on the minimum allocation of resources required to execute client jobs before their deadlines; that is, to avoid inaccurate run-time estimates of clients and thus mitigate SLA violations. The scheduler policy adopts a job rejection strategy in two different scenarios. A job is rejected when its QoS obligations cannot be met, or when another higher priority job arrives to the system that negatively impacts SLA obligations of both jobs. However, such rejection policies would incur harsh SLA violation penalties on the client and service provider.

In addition, Hoang \emph{et al}.~\cite{SAR2014} present a Soft Advance Reservation (SAR) method to meet SLA requirements and tackle error-prone estimates on job executions provided by the clients. Generally speaking, an over-sourced environment would reduce the likelihood of SLA violations and thus dissatisfied clients, however it would be significantly costly to acquire and operate. In contrast, the cloud service provider may allocate a small number of resources to reduce the operational cost, but with the expense of rejecting or discarding jobs that the provider would not meet their QoS expectations.

The meta-heuristic approaches are also presented to tackle scheduling problems in cloud computing environments~\cite{On_maximizing_2001, multi-dim2011, rahhali2018hybrid}. Such approaches are adopted to efficiently solve NP-hard computational-expensive problems, however the approaches deliver a near-optimal performance in a timely manner and potentially reduce the running time of the scheduling algorithms. Goudarzi \emph{et al}.~\cite{Maximizing_2011} present a heuristic-based allocation method to meet client SLAs and maximize the profit of the service provider in a data center of multiple clusters. However, each cluster adopts a centralized dispatcher associated with multiple resources comprising together a single-tier environment.

Zhang \emph{et al}.~\cite{SLA_Based_Profit_2004} propose a meta-heuristic scheduling algorithm that provides near-optimal resource configurations so as to maximize the profit and minimize the response time of jobs, in a centralized single-tier environment. Also, Zuo \emph{et al}.~\cite{multiObj2015} present an Ant Colony Optimization based scheduling method that finds a balance between the system performance represented by the makespan of jobs and the budget cost on the client. The former meta-heuristic approaches also tackle the problem in a single-tier environment and typically aim at optimizing the performance of schedules \emph{locally} at the individual resource level of the tier, similar to Min-Min and Max-Min based approaches. However, they do not support the complexity and obligations of the multi-tier environment, therefore do not produce job schedules that are optimized at the multi-tier level and thus would not accurately mitigate SLA penalties.

As a general observation, current scheduling approaches in cloud computing fail to contemplate the impact of schedules optimized in a given tier on the performance of schedules on the subsequent tiers. Such approaches do not effectively tackle dependencies between tiers of the multi-tier cloud environment. Instead, the approaches evaluate the optimality of schedules at the individual resource level of the single-tier environment, therefore SLA violation penalties in a tier would typically shift to and escalate in subsequent tiers leading to a potential increase in the likelihood of dissatisfied clients.

Furthermore, the reality is that clients of cloud computing have different computational demands and strict QoS expectations. Client jobs demand for services from multiple cloud resources characterized by multiple tiers of execution. Such jobs sometimes are delay-intolerant and tightly coupled with client satisfactions, and thus cannot afford SLA violation penalties. Workload variations occur within a short period of time and are not easily predictable, thus causing execution difficulties on the cloud service provider to fulfill such expectations and deliver optimal performance. Due to resource limitations and the complexity incurred from the multi-tier dependencies, formulating optimal schedules to satisfy various QoS obligations of client demands at the multi-tier level while maintaining high system performance is not a trivial task.

In this paper, a penalty-oriented approach is proposed to influence scheduling in the multi-tier cloud environment. The proposed approach contemplates tier dependencies to produce minimum-penalty schedules at the multi-tier level. The SLA violation penalties of job schedules in a tier are to be alleviated when jobs progress through subsequent tiers, and accordingly the performance of such schedules is optimized \emph{globally} at the multi-tier level. Since the problem is NP-hard, a biologically inspired meta-heuristic approach along with system virtualized and segmented queue abstractions are proposed to efficiently seek (near-)optimal schedules in a reasonable time.

\section{Penalty-Oriented Multi-Tier SLA Centric Scheduling of Cloud Jobs}
\label{sec:probFormal}

A multi-tier cloud computing environment consisting of $N$ sequential tiers is considered:
\begin{equation}
T = \left \{T\!_1,T\!_2,T\!_3,...,T\!_N  \right \}
\end{equation}

Each tier $T\!_j$ employs a set of identical computing resources $R_j$:
\begin{equation}
R_j = \left \{R_{j,1},R_{j,2},R_{j,3},...,R_{j,M}  \right \}
\end{equation}

Each resource $R_{j,k}$ employs a queue $Q_{j,k}$ that holds jobs waiting for execution by the resource. Jobs with different resource computational requirements and QoS obligations are submitted to the environment. It is assumed that these jobs are submitted by different clients and hence are governed by various SLA's. Jobs arrive at the environment in streams. A stream $S$ is a set of jobs:
\begin{equation}
S = \left \{ J_1,J_2,J_3,...,J_l  \right \}
\end{equation}

The index of each job $J_i$ signifies its arrival ordering at the environment. For example, job $J_1$ arrives at the environment before job $J_2$. 
Jobs arrive in random manner. Job $J_i$ arrives at tier $T\!_j$ at time $A_{i,j}$ via the queue of the job dispatcher $J\!D_j$ of the tier. It has a prescribed execution time $\mathcal{E}_{i,j}$ at each tier. Each job has a service deadline which in turn stipulates a target completion time $\mathcal{C}_i^{(t)}$ for the job $J_i$ in the multi-tier environment.
\begin{equation}
\label{equ:Ji}
    J_i = \left \{A_{i,j}, \mathcal{E}_{i,j}, \mathcal{C}_i^{(t)} \right \},\;\;\; \forall\; T\!\!_j\!\in\!T
\end{equation}

\begin{figure}[!t]
\centering
	  \includegraphics[width=\textwidth]{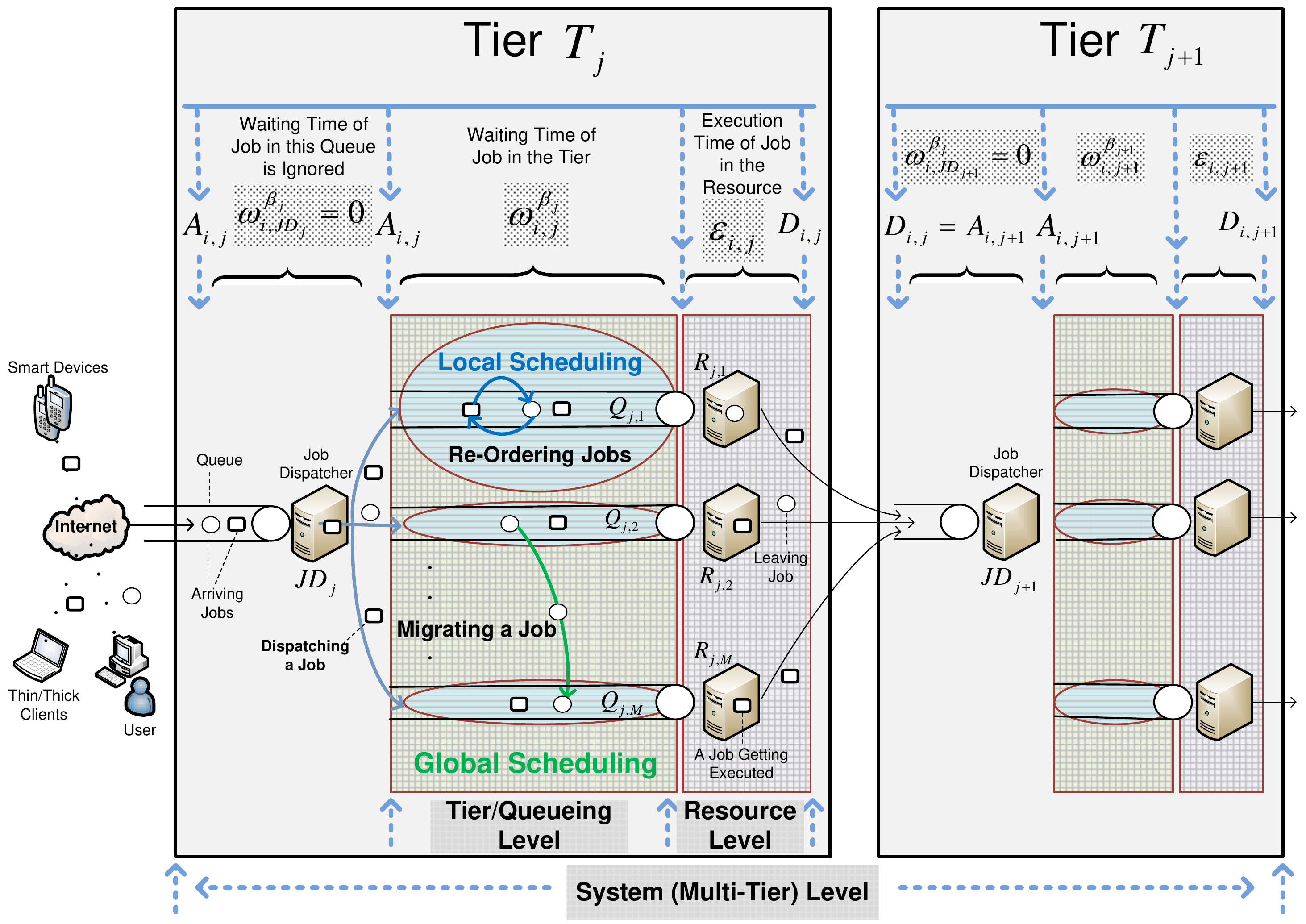}
	  \caption{System Model of the Multi-Tier Environment}
      \label{fig:Paper2_SystemModel1_1}
\end{figure}

Jobs submitted to tier $T\!_j$ are queued for execution based on an ordering $\beta_j$. As shown in Figure~\ref{fig:Paper2_SystemModel1_1}, each tier $T\!_j$ of the environment consists of a set of resources $R_j$. Each resource $R_{j,k}$ has a queue $Q_{j,k}$ to hold jobs assigned to it. For instance, resource $R_{j,1}$ of tier $T_j$ is associated with queue $Q_{j,1}$, which consists of $4$ jobs ($J_6$, $J_7$, $J_8$, and $J_{10}$) waiting for execution. A virtual-queue is a cascade of all queues of the tier as shown in Figure~\ref{fig:systemVirtualQueue}. The total execution time $\mathcal{E\!T}\!\!_{i}$ of each job $J_i$ is as follows:
\begin{equation}
\label{equ:Ei}
    \mathcal{E\!T}\!\!_i = \sum_{j=1}^{N} \mathcal{E}_{i,j}
\end{equation}

The target completion time $\mathcal{C}_i^{(t)}$ of job $J_i$ represents an explicit QoS obligation on the service provider to complete the execution of the job. Thus, the $\mathcal{C}_i^{(t)}$ incurs a service deadline $\mathcal{D\!L}_i$ for the job in the environment. The service deadline $\mathcal{D\!L}_i$ is higher than the total prescribed execution time $\mathcal{E\!T}\!\!_i$ and incurs a total waiting time allowance $\omega\!\mathcal{A\!L}_i$ for job $J_i$ in the environment.
\begin{equation}
\label{equ:DLi}
\begin{split}
   \mathcal{D\!L}_i & = \mathcal{C}_i^{(t)} - A_{i,j} \\
                    & = \mathcal{E\!T}\!\!_i + \omega\!\mathcal{A\!L}_i
\end{split}
\end{equation}

Each job $J_i$ has a response time $\mathcal{R\!T}_{\!i}^\beta$ that is a function of the total execution time $\mathcal{E\!T}\!\!_i$ and the total waiting time $\omega\!\mathcal{T}_i^\beta$.
\begin{equation}
\label{equ:Zi}
    \mathcal{R\!T}_{\!\!\!i}^\beta = \sum_{j=1}^{N} (\mathcal{E}_{i,j} + \omega_{i,j}^{\beta_j}) = \mathcal{E\!T}\!\!_i + \omega\!\mathcal{T}_i^\beta
\end{equation}
where $\omega_{i,j}^{\beta_j}$ represents the waiting time of job $J_i$ at tier $T_j$; $\beta_j$ is the ordering that governs the order of execution of jobs at tier $T_j$. The $\omega\!\mathcal{T}_i^\beta$ represents the total waiting time of job $J_i$ spends waiting for its turn to be executed at all tiers $T$ of the environment, according to the ordering $\beta$. Each job $J_i$ has a departure time $D_{i,j}$ from tier $T_j$, which will be the arrival time $A_{i,j+1}$ of the job to the next tier $T_{j+1}$.
\begin{equation}
\label{equ:beta}
    \beta = \bigcup_{j=1}^{N} \beta_j
\end{equation}

As such, the time difference between the response time $\mathcal{R\!T}_{\!\!\!i}^\beta$ and the service deadline $\mathcal{D\!L}_i$ represents the service-level violation time $\alpha_i^\beta$ of job $J_i$, according to the ordering $\beta$ of jobs in tiers $T$ of the environment.
\begin{equation}
\label{equ:alpha_i}
(\mathcal{R\!T}_{\!\!\!i}^\beta - \mathcal{D\!L}_i) =
\begin{cases}
     \alpha_i^\beta > 0, \;\; \text{The client is not satisfied}  \\
     \alpha_i^\beta \leq 0, \;\; \text{The client is satisfied}
\end{cases}
\end{equation}

However, the execution time $\mathcal{E}_{i,j}$ of job $J_i$ at tier $T\!_j$ is pre-defined in advance. Therefore, the resource capabilities of each tier $T_j$ are not considered and, thus, the total execution time $\mathcal{E\!T}\!\!_i$ of job $J_i$ is constant. Instead, the primary concern is on the queueing-level of the environment represented by the total waiting time $\omega\!\mathcal{T}_i^{\beta}$ of job $J_i$ at all tiers $T$ according to the ordering $\beta$.

Accordingly, the service-level violation time $\alpha_i^\beta$ of job $J_i$ in the environment is subject to an SLA that stipulates an exponential penalty curve $\varrho_i$:
\begin{equation}
\label{equ:rhoi2}
\begin{split}
    \varrho_i & = \chi * (   1- \text{e}^{- \nu(\mathcal{R\!T}_{\!\!i}^\beta - \mathcal{D\!L}_i) }   )\\
              & = \chi * (   1- \text{e}^{- \nu(\omega\!\mathcal{T}_i^\beta - \omega\!\mathcal{A\!L}_i) }   ) \\
              & = \chi * (   1- \text{e}^{- \nu(\alpha_i^\beta) }   )
\end{split}
\end{equation}
where $\chi$ is a monitory cost factor and $\nu$ is an arbitrary scaling factor. As such, the total penalty cost of stream $l$ across all tiers is given by $\varphi$:
\begin{equation}
\label{equ:Phi}
\begin{split}
   \varphi & = \sum_{i=1}^{l} \varrho_i
\end{split}
\end{equation}

\subsection{Multi-Tier Waiting Time Allowance $\omega\!\mathcal{A\!L}_i$ Formulation}

The performance of job schedules is formulated with respect to the multi-tier waiting time allowance $\omega\!\mathcal{A\!L}_i$ of each job $J_i$. Accordingly, the SLA violation penalty is evaluated at the multi-tier level of the environment. The objective is to seek job schedules in tiers of the environment such that the total SLA violation penalty of jobs would be minimized \emph{globally} at the multi-tier level of the environment.

The total waiting time $\omega\!\mathcal{T}^\beta_i$ of job $J_i$ currently waiting in tier $T\!_p$, where $\nolinebreak{p\!<\!N}$, is not totally known because the job has not yet completely finished execution from the multi-tier environment. Therefore, the job's $\omega\!\mathcal{T}^\beta_i$ at tier $T\!_p$ is estimated and, thus, represented by $\omega\mathcal{C\!X}_{i,p}^\beta$ according to the scheduling order $\beta$ of jobs. As such, the job's service-level violation time $\alpha_i^{\beta}$ at tier $T\!_p$ would be represented by the expected waiting time $\omega\mathcal{C\!X}_{i,p}^\beta$ of job $J_i$ in the current tier $T\!_p$ and the waiting time allowance $\omega\!\mathcal{A\!L}_i$ incurred from the job's service deadline $\mathcal{D\!L}_i$ at the multi-tier level of the environment.
\begin{equation}
\label{equ:alpha_i}
\begin{split}
   \alpha_i^\beta & = \omega\mathcal{C\!X}_{i,p}^\beta - \omega\!\mathcal{A\!L}_i
\end{split}
\end{equation}
where the expected waiting time $\omega\mathcal{C\!X}_{i,p}^\beta$ of job $J_i$ at tier $T\!_p$ incurs the total waiting time $\omega\!\mathcal{T}_i^\beta$ of job $J_i$ at the multi-tier level.
\begin{equation}
\label{equ:wEip}
\begin{split}
   \omega\mathcal{C\!X}_{i,p}^\beta & = \sum_{j=1}^{(p-1)}(\omega_{i,j}^{\beta_j}) + \omega\!E\!\mathcal{L}_{i,p} + \omega\!\mathcal{R\!M}_{i,p}^{\beta_p}
\end{split}
\end{equation}
where $\nolinebreak{\omega_{i,j}^{\beta_j}(\forall j\leq(p-1))}$ represents the waiting time of job $J_i$ in each tier $T\!_j$ in which the job has completed the execution in, $\nolinebreak{\omega\!E\!\mathcal{L}_{i,p}}$ represents the elapsed waiting time of job $J_i$ in the tier $T\!_p$ where the job currently resides, and $\nolinebreak{\omega\!\mathcal{R\!M}_{i,p}^{\beta_p}}$ represents the remaining waiting time of job $J_i$ according to the scheduling order $\beta_p$ of jobs in the current holding tier $T\!_p$.
\begin{equation}
\label{equ:betaj}
    \beta_j = \bigcup_{k=1}^{M_k} \text{I}(Q_{j,k}),\;\;\;\; \forall j\!\in\![1,N]
\end{equation}
\begin{equation}
\label{equ:wijbetaj}
    \omega\!\mathcal{R\!M}_{i,j}^{\beta_j} = \sum_{h\in \text{I}(Q_{j,k}),\;h\;\text{precedes job}\;J_i}^{\forall} \mathcal{E}_{h,j},\;\;\;\; \forall j\!\in\![1,N]
\end{equation}
where $\text{I}(Q_{j,k})$ represents indices of jobs in $Q_{j,k}$. For instance, $\nolinebreak{\text{I}(Q_{1,2})=\{3,5,2,7\}}$ signifies that jobs $J_3$, $J_5$, $J_2$, and $J_7$ are queued in $Q_{1,2}$ such that job $J_3$ precedes job $J_5$, which in turn precedes job $J_2$, and so on. However, the elapsed waiting time $\omega\!E\!\mathcal{L}_{i,j}$ affects the execution priority of the job. The higher the time of $\omega\!E\!\mathcal{L}_{i,j}$ of job $J_i$ in the tier $T\!_j$ the lower the remained allowed time of $\omega\!\mathcal{A\!L}_i$ of job $J_i$ at the multi-tier level, thus, the higher the execution priority of job $J_i$ in the resource.

The objective is to find scheduling orders $\nolinebreak{\beta=(\beta_1, \beta_2, \beta_3, \ldots, \beta_N)}$ for jobs of each tier $T\!_j$ such that the stream's total penalty cost $\varphi$ is minimal:
\begin{equation}
\label{equ:min_1}
\begin{split}
     \underset{\beta}{\text{minimize}} \; (\varphi) & \equiv \underset{\beta}{\text{minimize}} \;
     \big( \sum_{i=1}^{l}\sum_{p=1}^{N} (\omega\mathcal{C\!X}_{i,p}^\beta - \omega\!\mathcal{A\!L}_i) \big)
\end{split}
\end{equation}

\subsection{Differentiated Waiting Time Allowance $\omega\!\mathcal{P\!T}\!\!_{i,j}$ Formulation}

The performance of job schedules is formulated with respect to a differentiated waiting time $\omega\!\mathcal{P\!T}\!\!_{i,j}$ of the job $J_i$ at each tier $T\!_j$. The $\omega\!\mathcal{P\!T}\!\!_{i,j}$ is derived from the multi-tier waiting time allowance $\omega\!\mathcal{A\!L}_i$ of job $J_i$, with respect to the execution time $\mathcal{E}_{i,j}$ of the job $J_i$ at the tier level relative to the job's total execution time $\mathcal{E\!T}\!\!_i$ at the multi-tier level of the environment.
\begin{equation}
\label{equ:wPT}
\begin{split}
     \omega\!\mathcal{P\!T}\!_{i,j} = \omega\!\mathcal{A\!L}_i * \frac{\mathcal{E}_{i,j}}{\mathcal{E\!T}\!\!_i}
\end{split}
\end{equation}

In this case, the higher the execution time $\mathcal{E}_{i,j}$ of job $J_i$ in tier $T\!_j$, the higher the job's differentiated waiting time allowance $\omega\!\mathcal{P\!T}\!_{i,j}$ in the tier $T\!_j$. Accordingly, the SLA violation penalty is evaluated at the multi-tier level with respect to the $\omega\!\mathcal{P\!T}\!\!_{i,j}$ of each job $J_i$.

The waiting time $\omega_{i,j}^{\beta_j}$ of job $J_i$ at tier $T\!_j$ would not be totally known until the job completely finishes the execution from the tier, however, it can be estimated by $\omega\!\mathcal{P\!X}_{i,j}^{\beta_j}$ according to the current scheduling order $\beta_j$ of jobs in the tier $T\!_j$. As such, the service-level violation time $\alpha\!\mathcal{T}_{i,j}^{\beta_j}$ of job $J_i$ in the tier $T\!_j$ according to the scheduling order $\beta_j$ of jobs would be represented by the expected waiting time $\omega\!\mathcal{P\!X}_{i,j}^{\beta_j}$ and the differentiated waiting time allowance $\omega\!\mathcal{P\!T}\!\!_{i,j}$, of the job in the tier $T\!_j$.
\begin{equation}
\label{equ:alphaT_ij}
\begin{split}
   \alpha\!\mathcal{T}_{i,j}^{\beta_j} & = \omega\!\mathcal{P\!X}_{i,j}^{\beta_j} - \omega\!\mathcal{P\!T}\!\!_{i,j}
\end{split}
\end{equation}
\begin{equation}
\label{equ:alphaT_ij}
\begin{split}
   \alpha_{i}^{\beta} & = \sum_{j=1}^{N} \alpha\!\mathcal{T}_{i,j}^{\beta_j}
\end{split}
\end{equation}
where $\alpha_{i}^{\beta}$ is the total service-level violation time of the job $J_i$ at all tiers of the environment according to the scheduling order $\beta$. The expected waiting time $\omega\!\mathcal{P\!X}_{i,j}^{\beta_j}$ incurs the actual waiting time $\omega_{i,j}^{\beta_j}$ of job $J_i$ in tier $T\!_j$, and thus depends on the elapsed waiting time $\nolinebreak{\omega\!E\!\mathcal{L}_{i,j}}$ and the remaining waiting time $\nolinebreak{\omega\!\mathcal{R\!M}_{i,j}^{\beta_j}}$ of the job $J_i$ according to the scheduling order $\beta_j$ of jobs in the current holding tier $T\!_j$.
\begin{equation}
\label{equ:wCXij}
\begin{split}
   \omega\!\mathcal{P\!X}_{i,j}^{\beta_j} & = \omega\!E\!\mathcal{L}_{i,j} + \omega\!\mathcal{R\!M}_{i,j}^{\beta_j}
\end{split}
\end{equation}

The elapsed waiting time parameter $\omega\!E\!\mathcal{L}_{i,j}$ of job $J_i$ in tier $T\!_j$ affects the job's execution priority in the resource. The higher the time of $\omega\!E\!\mathcal{L}_{i,j}$, the lower the remained time of the differentiated waiting allowance $\omega\!\mathcal{P\!T}\!\!_{i,j}$ of job $J_i$ in the tier $T\!_j$, therefore the higher the execution priority of the job $J_i$ in the resource, so as to reduce the service-level violation time $\alpha\!\mathcal{T}_{i,j}^{\beta_j}$ of the job in the tier $T\!_j$ of the environment.

As such, the objective is to find scheduling orders $\nolinebreak{\beta=(\beta_1, \beta_2, \beta_3, \ldots, \beta_N)}$ for jobs of each tier $T\!_j$ such that the stream's total penalty cost $\varphi$ is minimal:
\begin{equation}
\label{equ:min_2}
\begin{split}
     \underset{\beta}{\text{minimize}} \; (\varphi) & \equiv \underset{\beta}{\text{minimize}} \; \big( \sum_{i=1}^{l} \sum_{j=1}^{N} (\omega\!\mathcal{P\!X}_{i,j}^{\beta_j} - \omega\!\mathcal{P\!T}\!_{i,j}) \big)
\end{split}
\end{equation}

\section{Multi-Tier-Based Minimum Penalty Scheduling: A Genetic Algorithm Formulation}
\label{sec:penaltyAppr}

This paper is concerned with the SLA-driven, penalty-based scheduling of jobs in a multi-tier cloud environment. The scheduling tackles tier dependencies by contemplating the impact of schedules optimized in a given tier on the performance of schedules in subsequent tiers. Thus, the potential of shifting and escalation of SLA violation penalties of schedules in a tier are mitigated when jobs progress through tiers of the environment. It is desired to produce job schedules that are penalty-minimum at the multi-tier level.

However, finding job schedules at the multi-tier level to minimize the SLA violation penalties is an NP problem. Jobs can be tightly coupled with the client experience and QoS obligations. Given the prohibitively large number of candidate schedules (permutations) of an excessive volume of critical jobs with their computational complexity in a multi-tier environment, it is never desirable to adopt the brute-force search strategy to seek minimum penalty schedules at the multi-tier level. The dimensionality of the search space of the multi-tier environment demands for an effective strategy that finds acceptable solutions. Thus, a meta-heuristic search strategy, such as Permutation Genetic Algorithms (PGA), is a viable option for efficiently exploring and exploiting the large space of scheduling permutations~\cite{GA3}. Genetic algorithms have been successfully adopted in various problem domains and shown less computational effort~\cite{GaTabu1}. They have undisputed success in yielding near optimal solutions for large scale problems, in reasonable time~\cite{heuristicPSO}.

Scheduling the client jobs entails two steps: (1) allocating/distributing the jobs among the different tier resources. Jobs that are allocated to a given resource are queued in the queue of that resource; (2) ordering the jobs in the queue of the resource such that their total SLA violation time is minimal. What makes the problem increasingly hard is the fact that jobs continue to arrive, while the prior jobs are waiting in their respective queues for execution. Thus, the scheduling process needs to respond to the job arrival dynamics to ensure that job execution at all tiers remains waiting-time optimal. To achieve this, job ordering in each queue should be treated as a continuous process. Furthermore, jobs should be migrated from one queue to another so as to ensure balanced job allocation and maximum resource utilization. Thus, the two operators are employed to construct optimal job schedules:
\begin{itemize}
  \item The \emph{reorder} operator is used to change the ordering of jobs in a given queue so as to find an ordering that minimizes the total SLA violation time of all jobs in the queue.
  \item The \emph{migrate} operator, in contrast, is used to exploit the benefits of moving jobs between the different resources of the tier so as to reduce the total SLA violation time at the multi-tier level. This process is adopted at each tier of the environment.
\end{itemize}
\begin{figure}[!t]
\captionsetup{justification=centering}
\centering
	  \includegraphics[width=\textwidth]{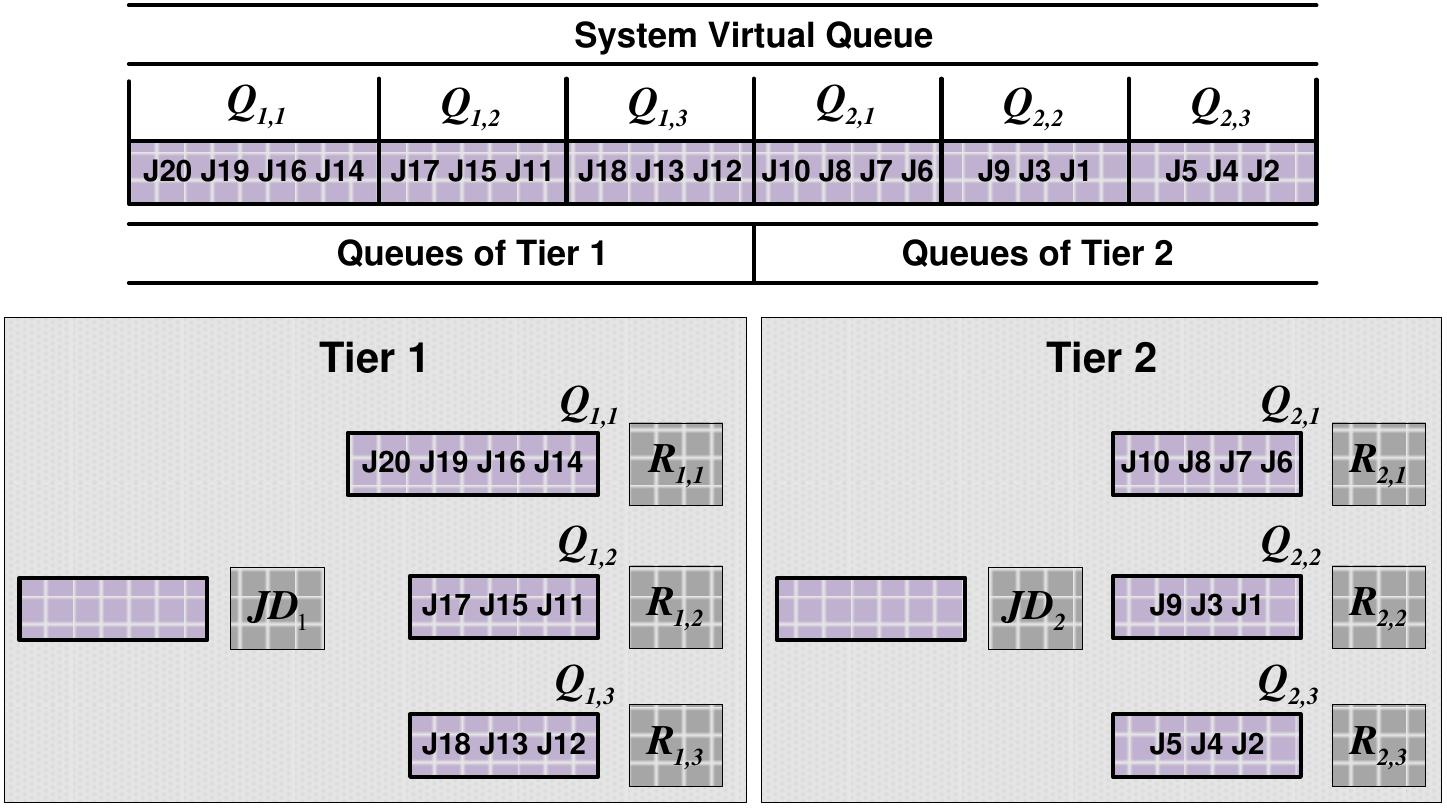}
	  \caption{The System Virtual Queue}
      \label{fig:systemVirtualQueue}
\end{figure}

However, implementing the \emph{reorder/migrate} operators in a PGA search strategy to create job schedules at the multi-tier level of the environment is not a trivial task. This implementation complexity can be relaxed by virtualizing queues of the tiers into one \emph{system virtual queue}. As shown in Figure~\ref{fig:systemVirtualQueue}, the system virtual queue is simply a cascade of the resource queues of the multi-tier environment.

\begin{figure*}[!t]
\captionsetup{justification=centering}
\centering
	  \includegraphics[width=\textwidth]{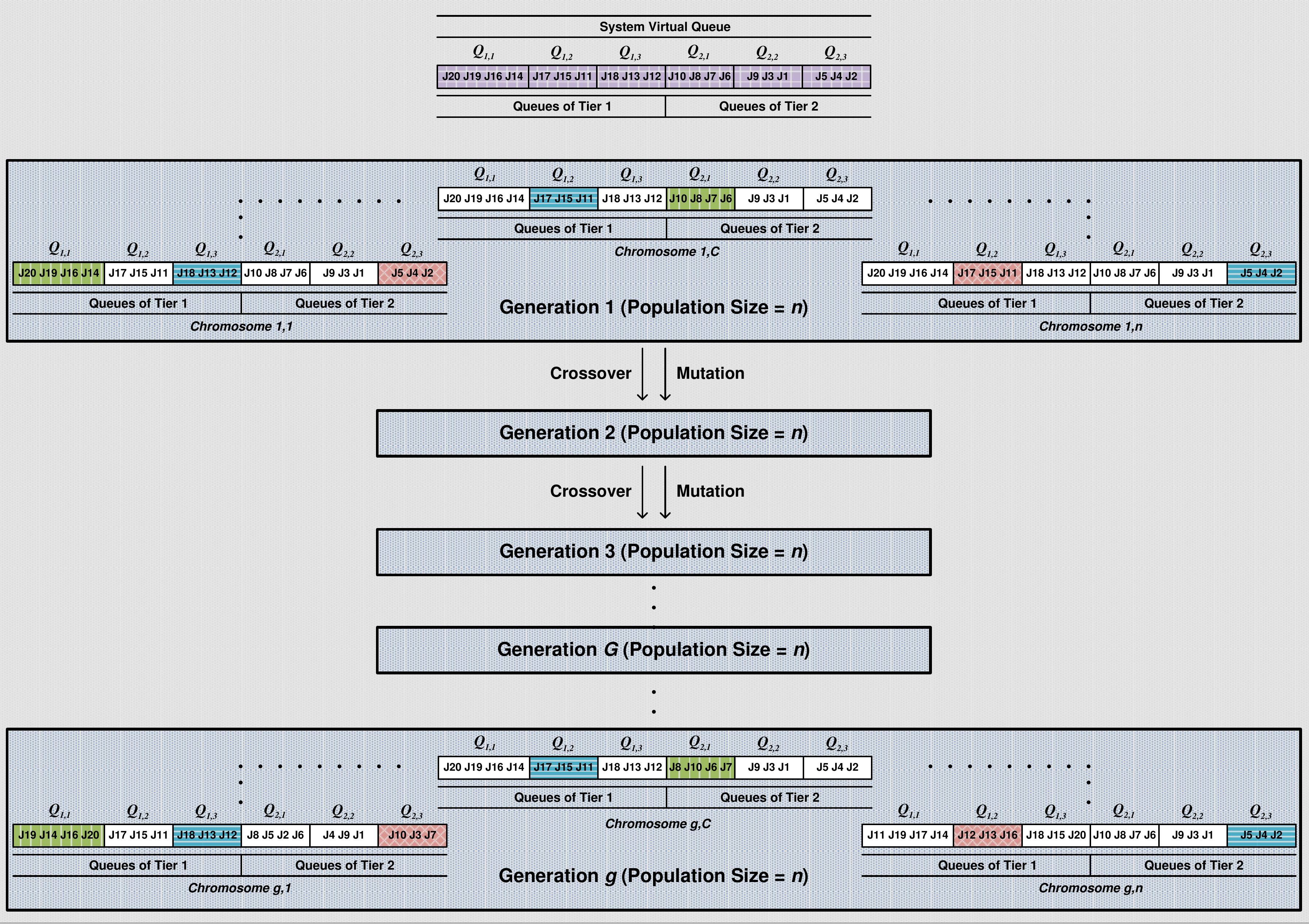}
	  \caption{A System Virtualized Queue Genetic Approach}
      \label{fig:GA}
\end{figure*}

In this way, the reorder/migrate operators running at the queue/tier level are converged into simply a reorder operator running at the multi-tier level. This system virtualization simplifies the PGA solution formulation toward finding schedules that are penalty-minimum at the multi-tier level. A consequence of this abstraction is the length of the permutation chromosome and the associated computational cost. This system virtual queue will serve as the chromosome of the solution that represents the scheduling of jobs on resource queues of tiers. An index of a job in this queue represents a gene. The ordering of jobs in a system virtual queue signifies the order at which the jobs in this queue are to be executed by the resource associated with that queue. Solution populations are created by permuting the entries of the system virtual queue, using the \emph{order} and \emph{migrate} operators. The system virtual queue in Figures~\ref{fig:systemVirtualQueue} and \ref{fig:GA} has six queues ($Q_{1,1}$, $Q_{1,2}$, $Q_{1,3}$, $Q_{2,1}$, $Q_{2,2}$, and $Q_{2,3}$) cascaded to construct one system virtual queue.

\subsection{Evaluation of Schedules}
\label{sec:step3}

The quality of a job schedule in a system virtual queue realization (chromosome) is assessed by a fitness evaluation function. For a chromosome $r$ in generation $G$, the fitness value $f_{r,G}$ is represented by the SLA violation cost of the schedule in the system virtual queue computed at the multi-tier level. Two different fitness evaluation functions are adopted in two different solutions:
\begin{equation}
\label{equ:fitness2}
f_{r,G} = \\
\begin{cases}
     \sum_{i=1}^{l} (\omega\mathcal{C\!X}_{i,p}^\beta - \omega\!\mathcal{A\!L}_i), \; \omega\!\mathcal{A\!L}_i~\text{based Scheduling}  \\
     \sum_{i=1}^{l} (\omega\!\mathcal{P\!X}_{i,j}^{\beta_j} - \omega\!\mathcal{P\!T}\!_{i,j}), \; \omega\!\mathcal{P\!T}\!_{i,j}~\text{based Scheduling}
\end{cases}
\end{equation}

In both scenarios, the SLA violation cost of job $J_i$ is represented by the job's waiting time (either $\omega\mathcal{C\!X}_{i,p}^\beta$ or $\omega\!\mathcal{P\!X}_{i,j}^{\beta_j}$) according to its scheduling order $\beta$ in the system virtual queue and the job's waiting allowance (either $\omega\!\mathcal{A\!L}_i$ or $\omega\!\mathcal{P\!T}\!_{i,j}$) incurred from the job's deadline $\mathcal{D\!L}_i$ at the multi-tier level.

The normalized fitness value $F_r$ of each schedule candidate is computed as follows:
\begin{equation}
  \label{equ:probabilityRouletteWheel}
  F_r  = \frac{f_{r,G}}{\sum_{C=1}^{n} (f_{C,G})}\;,\;\;\;r\!\in\!C
\end{equation}

Based on the normalized fitness values of the candidates, the Russian Roulette is used to select a set of schedule candidates to produce the next generation population, using the combination and mutation operators.

\subsection{Evolving the Scheduling Process}
\label{sec:step4}

The job schedule of the system virtual queue is evolved to produce a population of multiple system virtual queues, each of which represents a chromosome that holds a new scheduling order of jobs in resource queues of the multi-tier environment. The crossover and mutation genetic operators are applied on randomly selected system virtual queues from the current population to produce the new population. Such operators explore and exploit the search space of possible scheduling options without getting stuck in locally optimum solutions. The \emph{Single-Point} crossover and \emph{Insert} mutation operators are used; rates of these operators in each generation are set to be $0.1$ of the population size.

The evolution process of schedules of the system virtual queues along with the genetic operators are explained in Figure~\ref{fig:GA}. Each segment in the system virtual queue corresponds to an actual queue associated with a resource in the tier. In each generation, each segment is subject to one of the following states:
\begin{itemize}
  \item Maintain the same scheduling set and order of jobs held in the previous generation;
  \item Get a new scheduling order for the same set of jobs held in the previous generation;
  \item Get a different scheduling set and order of jobs.
\end{itemize}

For instance, queue $Q_{2,3}$ of \nolinebreak{\emph{Chromosome} ($1,\!n$)} in the first generation maintains exactly the same scheduling set and order of jobs in the final generation shown in queue $Q_{2,3}$ of \nolinebreak{\emph{Chromosome} ($g,\!n$)}. In contrast, queue $Q_{1,1}$ of \nolinebreak{\emph{Chromosome} ($1,\!1$)} in the first generation maintains the same scheduling set of jobs in the final generation, yet has got a new scheduling order of jobs as shown in queue $Q_{1,1}$ of \nolinebreak{\emph{Chromosome} ($g,\!1$)}. A similar observation is shown in queue $Q_{2,1}$ of \nolinebreak{\emph{Chromosomes} ($1,\!C$)} and ($g,\!C$) that has only got the scheduling order changed, however $Q_{2,2}$ and $Q_{2,3}$ of the same tier have got the same scheduling set and order of jobs held in the first generation. On the other side, some other queues would neither maintain the same scheduling set nor the same scheduling order of jobs in the last generation, such as queue $Q_{1,2}$ of \nolinebreak{\emph{Chromosomes} ($1,\!n$)} and ($g,\!n$). Thus, if \nolinebreak{\emph{Chromosome} ($g,\!1$)} is later selected as the best chromosome of the genetic solution, the state of the multi-tier environment is represented as follows:
\begin{itemize}
  \item Queues of resources $R_{1,2}$ and $R_{1,3}$ of the first tier $T\!_1$ would maintain the same schedules of jobs of the first generation.
  \item The queue of resource $R_{1,1}$ of the first tier $T\!_1$ would just get a new scheduling order of the same set of jobs held in the first generation.
  \item Queues of resources $R_{2,1}$, $R_{2,2}$, and $R_{2,3}$ of the second tier $T\!_2$ would hold totally new schedules of jobs.
\end{itemize}

\section{Experimental Work and Discussions on Results}
\label{sec:res}

The adopted cloud environment in this paper consists of two tiers, each of which has 3 computing resources. The jobs generated into the cloud environment are atomic and independent of each other. A job is first executed on one of the computing resources of the first tier and then moves for execution on one of the resources of the second tier. Each job is served by only one resource at a time, as the scheduling strategy is non-preemptive.

Jobs arrive at the first tier and are queued in the arrival queue (tier's dispatcher) of the environment. The arrival behaviour is modeled on a Poisson process. The running time of each job in a computing resource is assumed to be known in advance, generated with a rate $\nolinebreak{\mu\!=\!1}$ from the exponential distribution function $\nolinebreak{\text{exp}(\mu\!=\!1)}$~\cite{PerfEval_2016}. In each tier $T\!_j$, job migrations from a queue to another queue are permitted. The waiting time allowance $\omega\!\mathcal{A\!L}_i$ of each job $J_i$ is generated with respect to the job's total execution time $\mathcal{E\!T}\!\!_{i}$ at the multi-tier level of the environment as follows:
\begin{equation}
\label{equ:wALiGen}
\begin{split}
   \omega\!\mathcal{A\!L}_i = \mathcal{E\!T}\!\!_i * 20\%
\end{split}
\end{equation}

Accordingly, the differentiated waiting time allowance $\omega\!\mathcal{P\!T}\!\!_{i,j}$ of each job $J_i$ is generated using Equation~\ref{equ:wPT}.

\subsection{The Experimental Approach}

Two experiments are conducted, the system virtualized queue and segmented queue. To seek optimal schedules that produce minimum SLA penalty among all jobs at the multi tier level, the system virtual queue is employed and the multi-tier-driven genetic algorithm operates on all queues of the multi-tier environment simultaneously. The system virtual queue starts with an initial system-state and a QoS penalty that represent a schedule $\beta$ of jobs. The genetic solution finds an enhanced schedule that reduces the SLA penalty of the system-state at the multi-tier level, which in turn is translated into an enhanced schedule of jobs in the resource queues of tiers. In contrast, the segmented queue scheduling employs the genetic solution to seek an optimal schedule at the individual queue level of the tiers, in a reduced search space, such that the QoS penalty is reduced at the queue level of the tier and consequently at the multi-tier level. However, the penalty exponential scaling parameter $\nu$ is set to be $\nolinebreak{\nu\!=\!0.01}$. In both experiments, each population employs 10 chromosomes.

\subsection{QoS Penalty Scheduling Evaluation of the Waiting Time Allowance $\omega\!\mathcal{A\!L}_i$}

\begin{figure*}[!ht]
\captionsetup{justification=centering}
        \centering
        \begin{subfigure}{0.324\textwidth}
		        \includegraphics[width=\textwidth]
                {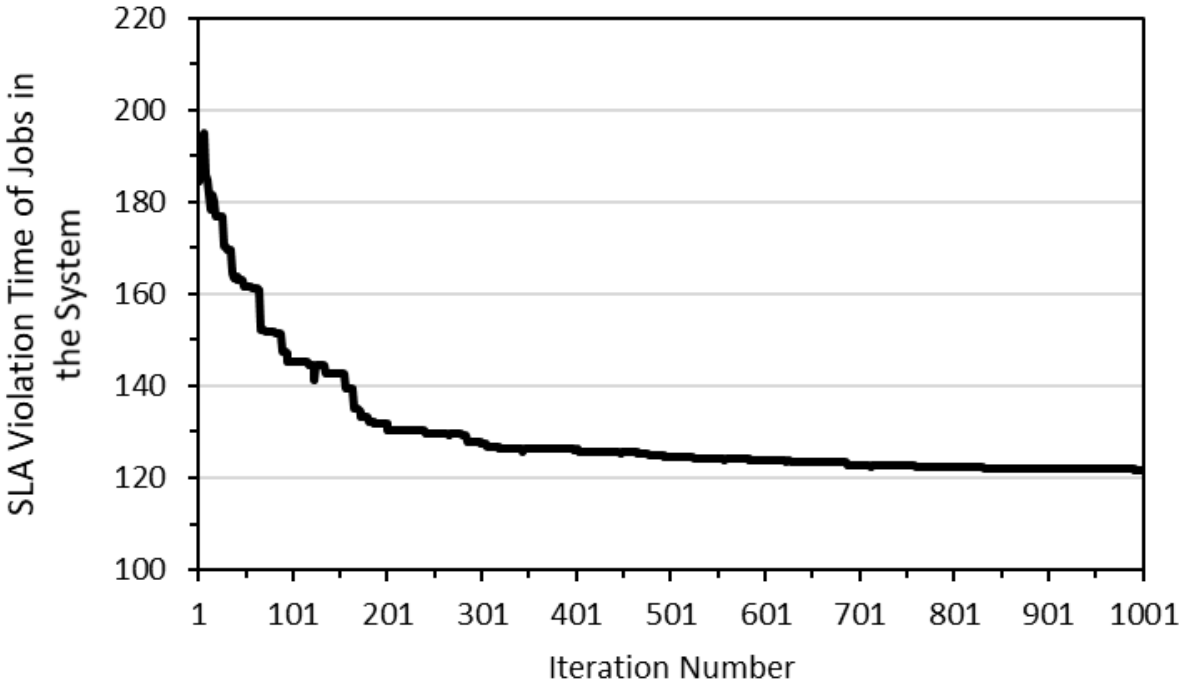}
	            \caption{System-Level (Total of 46 Jobs)}
	            \label{fig:SYSTEM_184_3927} 
        \end{subfigure}
        ~
        \begin{subfigure}{0.32\textwidth}
		        \includegraphics[width=\textwidth]
                {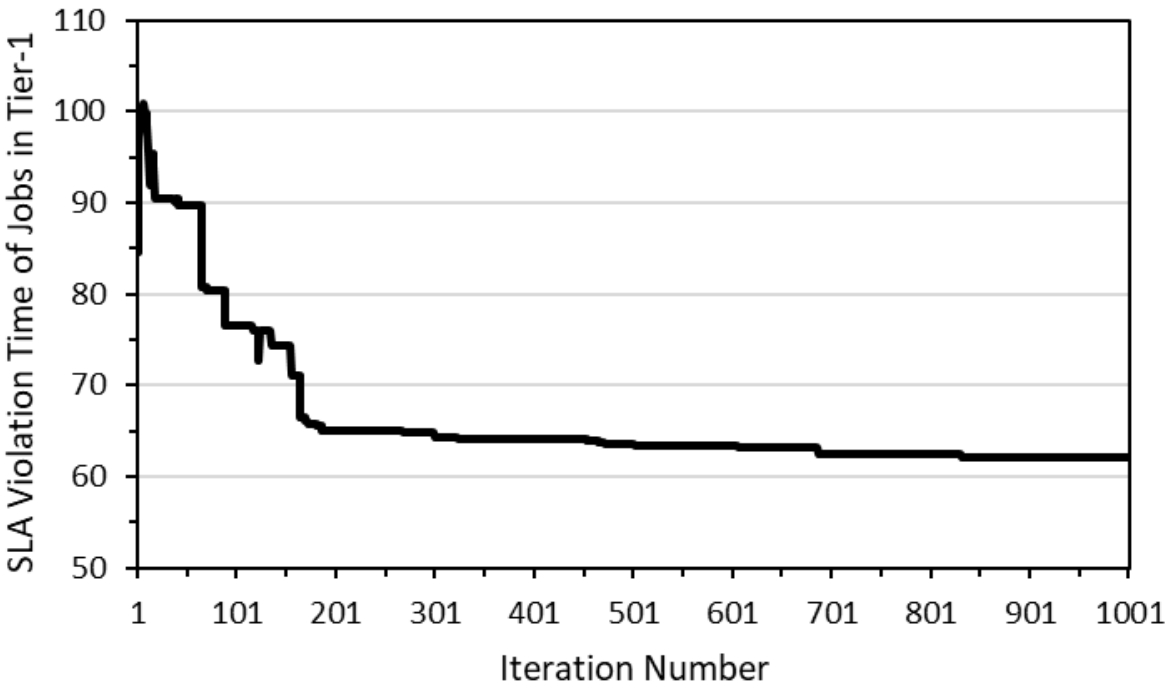}
	            \caption{Tier-1 (21 Jobs)}
	            \label{fig:Tier1SYSTEM_84_5951} 
        \end{subfigure}
        ~
        \begin{subfigure}{0.32\textwidth}
                \includegraphics[width=\textwidth]
                {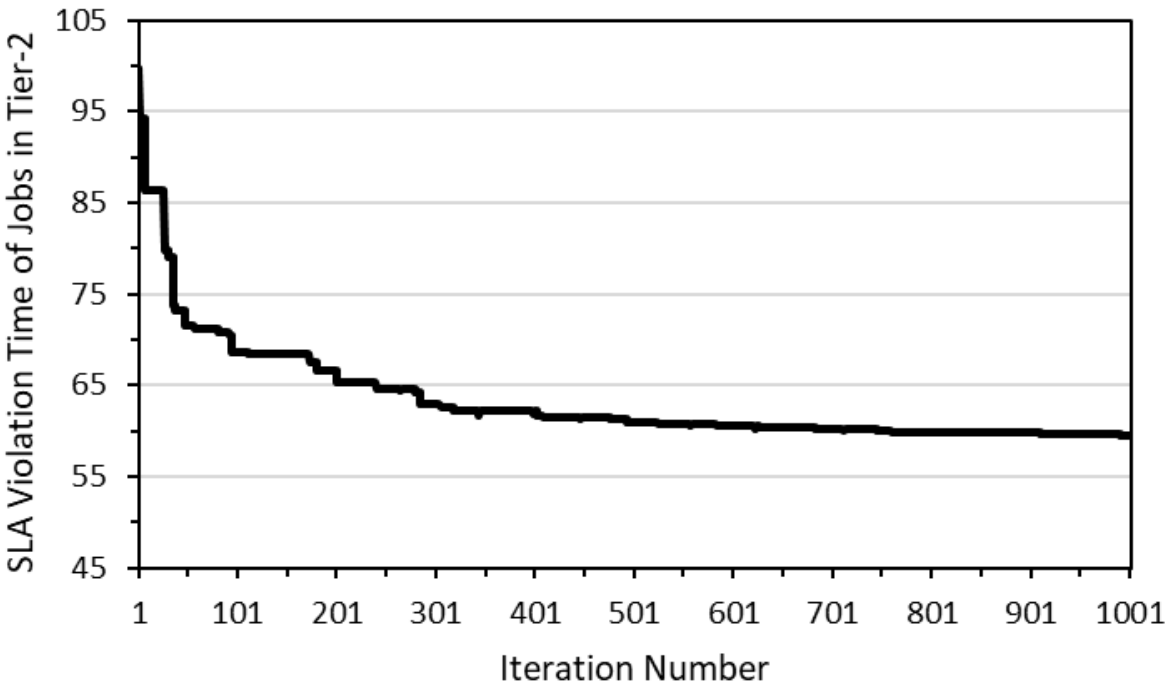}
	            \caption{Tier-2 (25 Jobs)}
	            \label{fig:Tier2SYSTEM_99_7976}
        \end{subfigure}

        \caption{System Virtualized Queue Scheduling with Respect to Multi-Tier $\omega\!\mathcal{A\!L}_i$}
        \label{fig:Case2_Tier_SYSTEM}
\end{figure*}

\begin{figure*}[!t]
\captionsetup{justification=centering}
        \centering
        \begin{subfigure}{0.32\textwidth}
		        \includegraphics[width=\textwidth]
                {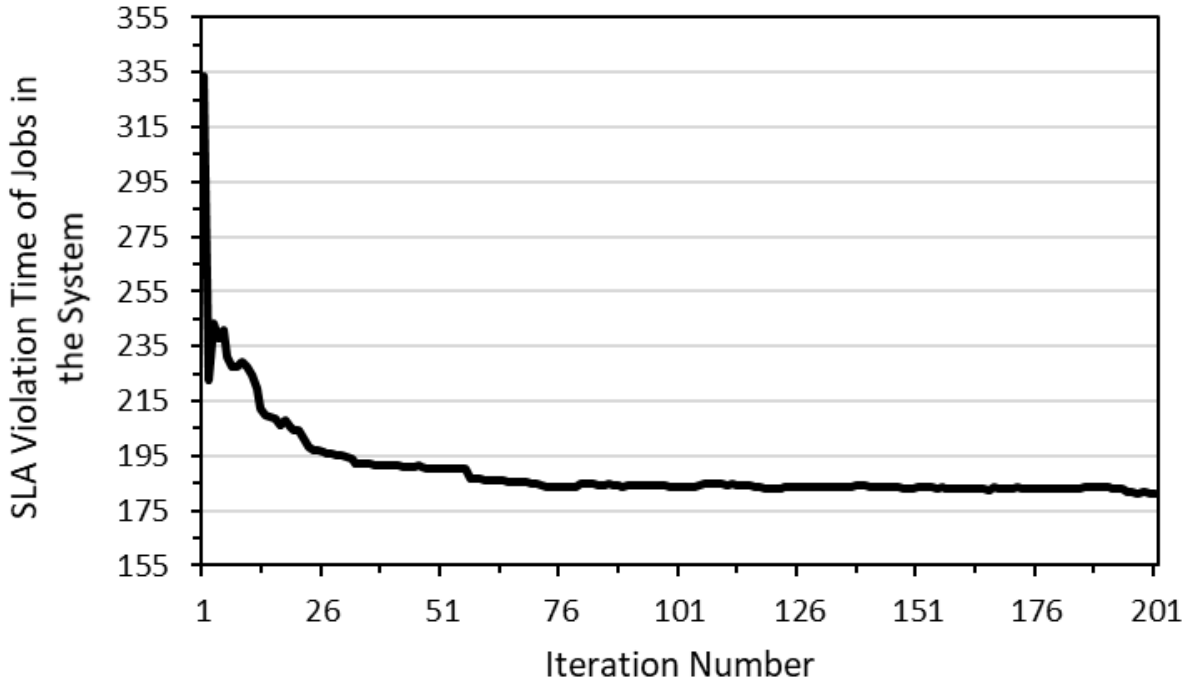}
	            \caption{System-Level (Total of 77 Jobs)}
	            \label{fig:SYSTEM_333_37}
        \end{subfigure}

        \begin{subfigure}{0.32\textwidth}
		        \includegraphics[width=\textwidth]
                {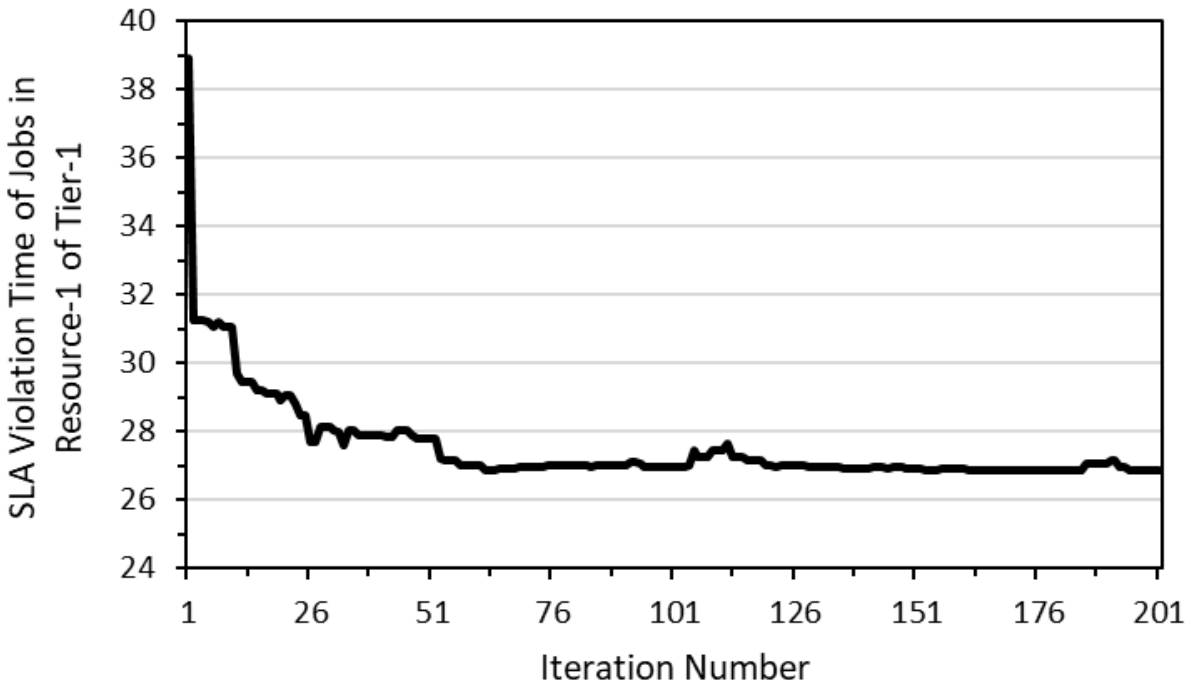}
	            \caption{Resource 1 of Tier 1 (Queue of 10 Jobs)}
	            \label{fig:Server1Tier1SYSTEM_38_9302}
        \end{subfigure}
        ~
        \begin{subfigure}{0.32\textwidth}
		        \includegraphics[width=\textwidth]
                {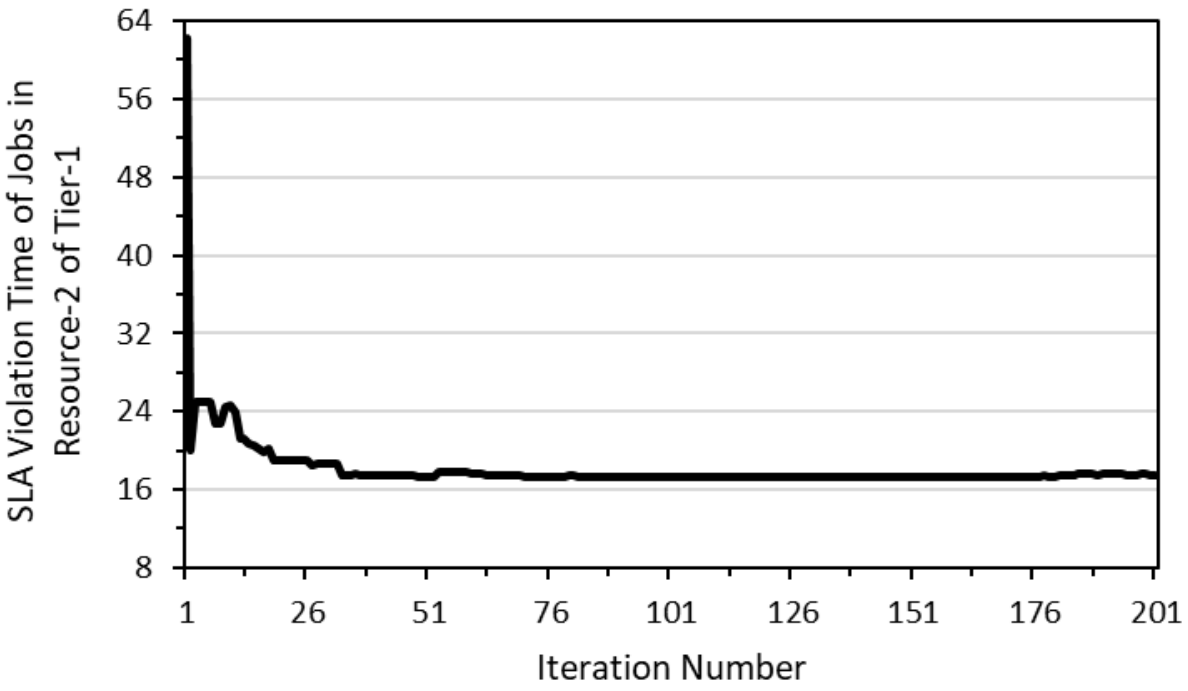}
	            \caption{Resource 2 of Tier 1 (Queue of 12 Jobs)}
	            \label{fig:Server2Tier1SYSTEM_317_3434}
        \end{subfigure}
        ~
        \begin{subfigure}{0.32\textwidth}
                \includegraphics[width=\textwidth]
                {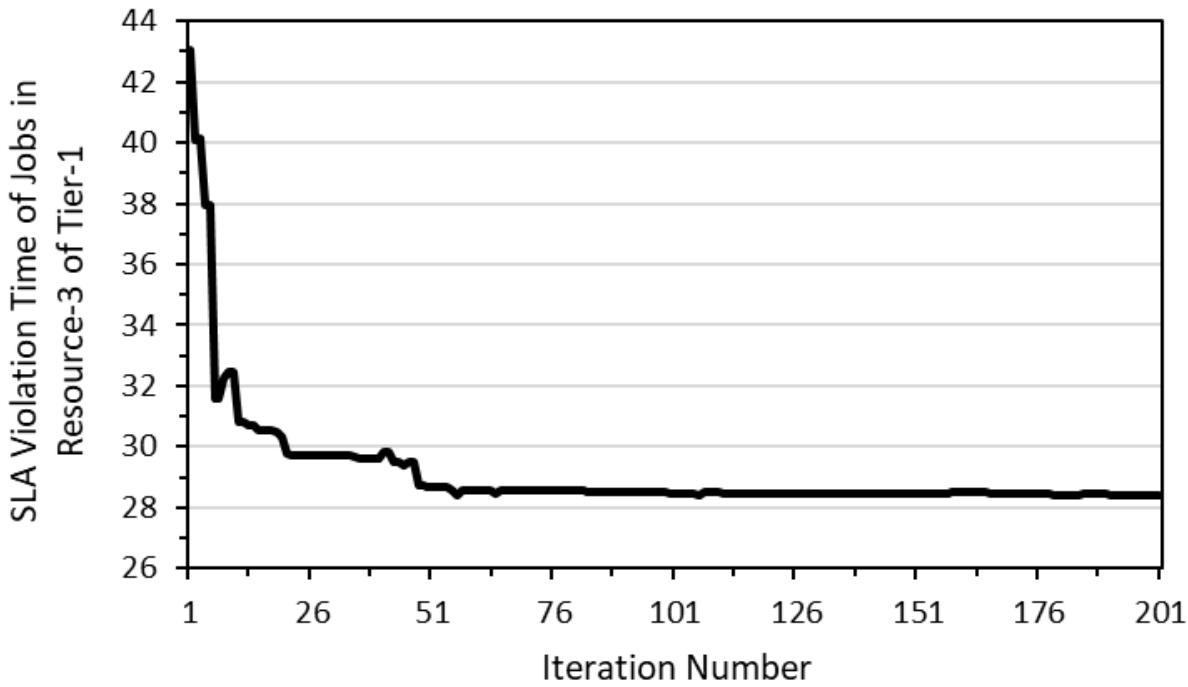}
	            \caption{Resource 3 of Tier 1 (Queue of 12 Jobs)}
	            \label{fig:Server3Tier1SYSTEM_43_0777}
        \end{subfigure}

        \begin{subfigure}{0.32\textwidth}
		        \includegraphics[width=\textwidth]
                {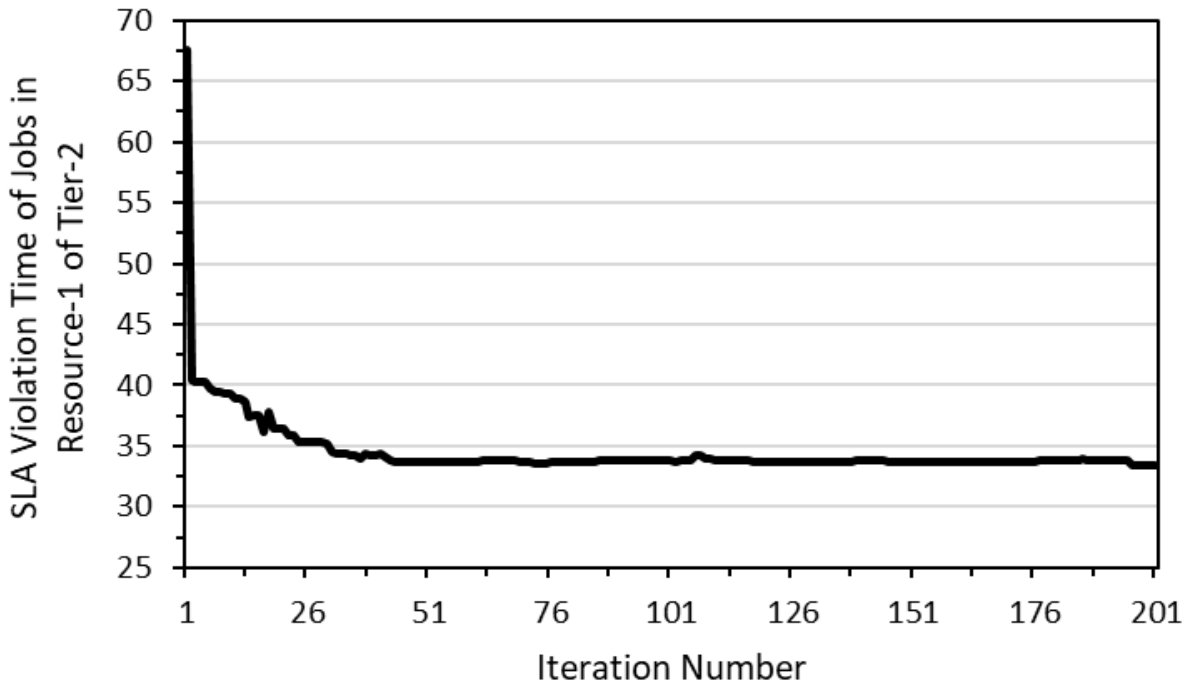}
	            \caption{Resource 1 of Tier 2 (Queue of 14 Jobs)}
	            \label{fig:Server1Tier2SYSTEM_67_5716}
        \end{subfigure}
        ~
        \begin{subfigure}{0.32\textwidth}
		        \includegraphics[width=\textwidth]
                {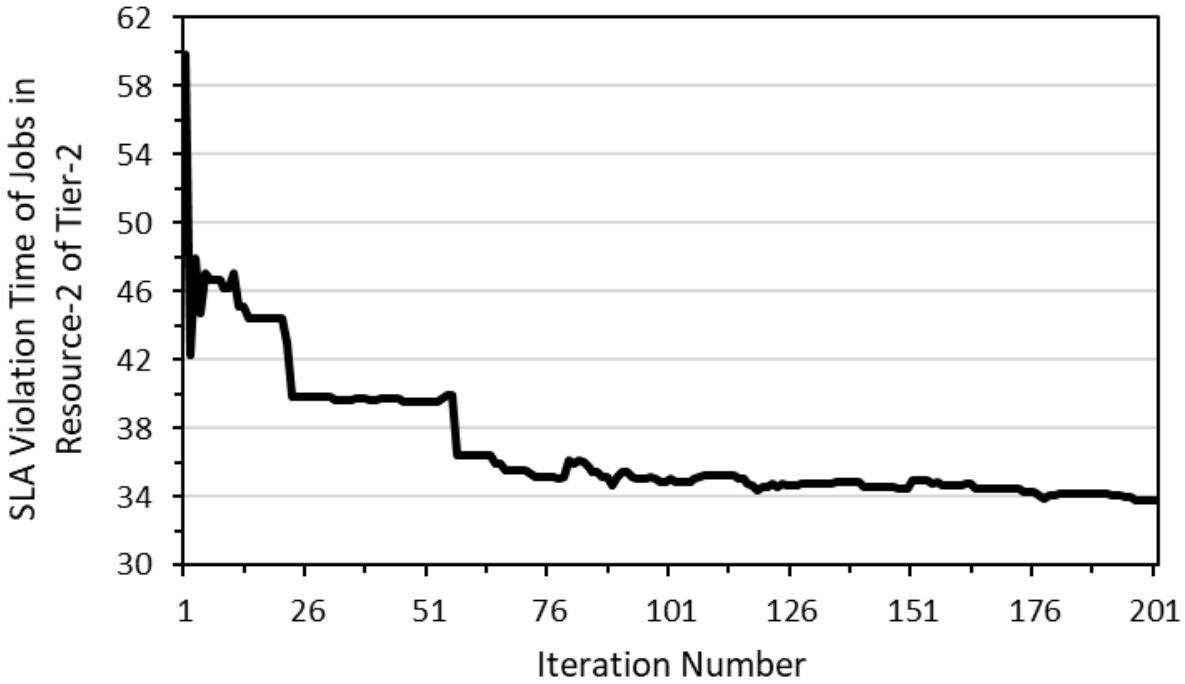}
	            \caption{Resource 2 of Tier 2 (Queue of 15 Jobs)}
	            \label{fig:Server2Tier2SYSTEM_59_8635}
        \end{subfigure}
        ~
        \begin{subfigure}{0.32\textwidth}
                \includegraphics[width=\textwidth]
                {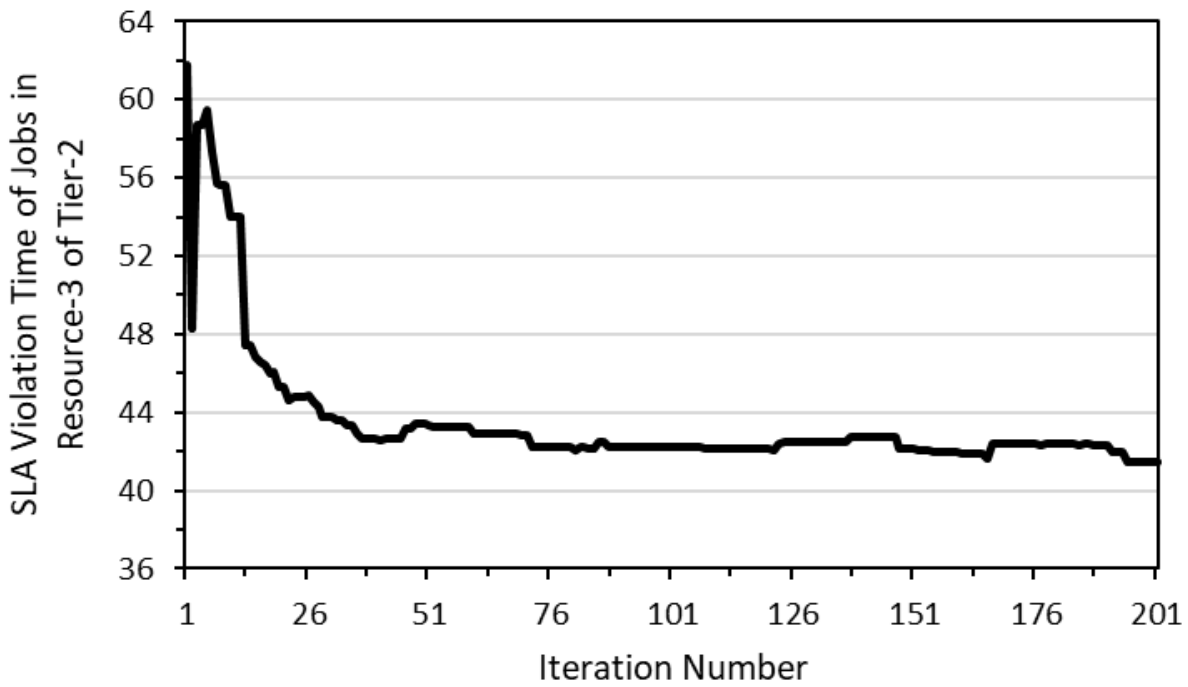}
	            \caption{Resource 3 of Tier 2 (Queue of 14 Jobs)}
	            \label{fig:Server3Tier2SYSTEM_61_7968}
        \end{subfigure}

        \caption{Segmented Queue Scheduling with Respect to Multi-Tier $\omega\!\mathcal{A\!L}_i$}
        \label{fig:Case2_Queue_SYSTEM}
\end{figure*}

The job schedules have been conducted according to the multi-tier waiting time allowance $\omega\!\mathcal{A\!L}_i$ of each job $J_i$. The service-level violation time of each job $J_i$ is measured at the multi-tier level with respect to the $\omega\!\mathcal{A\!L}_i$ of the job; accordingly, the SLA violation penalty payable by the service provider is quantified. The system virtualized queue and segmented queue genetic solutions are used to efficiently seek optimal job schedules. Overall, the scheduling approach has been proven to enhance the performance by producing optimal job schedules that reduce the total service-level violation time of jobs and their associated SLA penalty \emph{globally} at the multi-tier level of the environment (as shown in Figures~\ref{fig:Case2_Tier_SYSTEM} and \ref{fig:Case2_Queue_SYSTEM}, as well as Tables~1 and 2).

The scheduling approach along with the system virtualized queue genetic solution has been applied to seek an optimal scheduling of jobs. Figure~\ref{fig:Case2_Tier_SYSTEM} and Table~1 represent a state of a multi-tier environment that contains $46$ jobs; $21$ jobs are allocated to tier $T\!_1$ and $25$ jobs are allocated to tier $T\!_2$. At the start, the total service-level violation time of the initial scheduling order of the $46$ jobs on both tiers initiates with $184$ units of violation time (as shown in Figure~\ref{fig:SYSTEM_184_3927}). Then, the scheduling approach along with the system virtualized queue genetic setup has formed an enhanced schedule for the $46$ jobs on resource queues of both tiers, that optimizes the performance at the multi-tier level by $34\%$ to reach $121$ units of violation time. As a results, the SLA penalty payable by the service provider is also optimized by $24\%$, a reduction from $1.2$ for the initial schedule to $0.91$ for the enhanced schedule of the $46$ jobs (as shown in Table~1).

\begin{table*}[!ht]
\label{tab:Tier_SystemLevel}
\captionsetup{justification=centering}
\caption{System Virtualized Queue Scheduling with Respect to Multi-Tier $\omega\!\mathcal{A\!L}_i$}
\begin{center}\scalebox{0.82}{
\begin{threeparttable}
\begin{tabular}{c|c|cccc|cc}
\hline
\multirow{2}{*}{}
& \multirow{2}{*}{\textbf{\begin{tabular}[c]{@{}c@{}}Number \\of Jobs\end{tabular}}}\tnote{1}
& \multicolumn{2}{c}{\textbf{Initial}\tnote{2}} & \multicolumn{2}{c|}{\textbf{Enhanced}\tnote{3}}
& \multicolumn{2}{c}{\textbf{Improvement}}  \\ \cline{3-8}
& & \textbf{Violation}  & \textbf{Penalty} & \textbf{Violation} & \textbf{Penalty} & \textbf{Violation \%} & \textbf{Penalty \%} \\ \hline \hline

System-Level, Figure~\ref{fig:SYSTEM_184_3927}
& 46 & 184.39 & 1.2    & 121.69 & 0.91  & 34.01\%  & 24.17\%   \\ \hline
Tier-1, Figure~\ref{fig:Tier1SYSTEM_84_5951}
& 21 & 84.60  & 0.57   & 62.16  & 0.46  & 26.53\%  & 18.91\%   \\
Tier-2, Figure~\ref{fig:Tier2SYSTEM_99_7976}
& 25 & 99.80  & 0.63   & 59.53  & 0.45  & 40.35\%  & 28.95\%   \\ \hline

\end{tabular}
\begin{tablenotes}\scriptsize
\item[1] \textbf{Number of Jobs} represents the total number of jobs in queues of the tier/environment. For instance, the first entry (46 jobs) shows that the multi-tier environment contains 46 jobs in total. The second (21 jobs) and third (25 jobs) entries of the table mean that the 3 queues of tier-1 and tier-2 are allocated 21 and 25 jobs, respectively.
\item[2] \textbf{Initial Violation} represents the total SLA violation time of jobs according to their initial scheduling before using the system virtualized queue genetic solution.
\item[3] \textbf{Enhanced Violation} represents the total SLA violation time of jobs according to their final/enhanced scheduling found after using the system virtualized queue genetic solution.
\end{tablenotes}
\end{threeparttable}}
\end{center}
\end{table*}

The former enhancements achieved \emph{globally} at the multi-tier level of the environment would consequently optimize the performance of job schedules in each individual tier, thus, reduce the total service-level violation time and SLA penalty of the virtual-queue of each tier. For instance, the initial schedule of the virtual-queue ($25$ jobs) of tier $T\!_2$ shown in Figure~\ref{fig:Tier2SYSTEM_99_7976} began with $99.8$ units of violation time. Then, the performance has been optimized by $40\%$ to reach $59.5$ units of violation time for the enhanced schedule of jobs as a consequence of applying the scheduling approach along with the system virtualized queue genetic setup. As such, the total SLA penalty of jobs at tier $T\!_2$ has been reduced by $28.95\%$ (as shown in Table~1). Similarly, the results reported in Figure~\ref{fig:Tier1SYSTEM_84_5951} and Table~1 demonstrate the effectiveness of the system virtualized queue scheduling approach in reducing the total service-level violation time and penalty of the virtual-queue ($21$ jobs) of tier $T\!_1$ by $26.5\%$ and $18.9\%$, respectively.

\begin{table*}[!h]
\label{tab:Queue_SystemLevel}
\captionsetup{justification=centering}
\caption{Segmented Queue Scheduling with Respect to Multi-Tier $\omega\!\mathcal{A\!L}_i$}
\begin{center}\scalebox{0.8}{
\begin{threeparttable}
\begin{tabular}{c|c|cccc|cc}
\hline
\multirow{2}{*}{} & \multirow{2}{*}{\textbf{\begin{tabular}[c]{@{}c@{}}Number \\of Jobs\end{tabular}}}
& \multicolumn{2}{c}{\textbf{Initial}\tnote{4}} & \multicolumn{2}{c|}{\textbf{Enhanced}\tnote{5}}
& \multicolumn{2}{c}{\textbf{Improvement}}  \\ \cline{3-8}
& & \textbf{Violation}  & \textbf{Penalty} & \textbf{Violation} & \textbf{Penalty} & \textbf{Violation \%} & \textbf{Penalty \%} \\ \hline \hline

System-Level, Figure~\ref{fig:SYSTEM_333_37}
& 77 & 333.37 & 2.537 & 181.26 & 1.56 & 45.63\% & 38.51\%   \\ \hline

Resource-1 Tier-1, Figure~\ref{fig:Server1Tier1SYSTEM_38_9302}
& 10 & 62.13 & 0.463 & 17.34 & 0.16 & 72.09\% & 65.59\%  \\
Resource-2 Tier-1, Figure~\ref{fig:Server2Tier1SYSTEM_317_3434}
& 12 & 38.93 & 0.322 & 26.84 & 0.24 & 31.05\% & 27.00\%  \\
Resource-3 Tier-1, Figure~\ref{fig:Server3Tier1SYSTEM_43_0777}
& 12 & 43.08 & 0.350 & 28.41 & 0.25 & 34.06\% & 29.35\%  \\ \hline

Resource-1 Tier-2, Figure~\ref{fig:Server1Tier2SYSTEM_67_5716}
& 14 & 67.57 & 0.491 & 33.43  & 0.28 & 50.52\% & 42.15\%   \\
Resource-2 Tier-2, Figure~\ref{fig:Server2Tier2SYSTEM_59_8635}
& 15 & 59.86 & 0.450 & 33.77  & 0.29 & 43.58\% & 36.37\%   \\
Resource-3 Tier-2, Figure~\ref{fig:Server3Tier2SYSTEM_61_7968}
& 14 & 61.80 & 0.461 & 41.46  & 0.34 & 32.91\% & 26.37\%   \\ \hline

\end{tabular}
\begin{tablenotes}\scriptsize
\item[4] \textbf{Initial Violation} represents the total SLA violation time of jobs according to their initial scheduling before using the segmented queue genetic solution.
\item[5] \textbf{Enhanced Violation} represents the total SLA violation time of jobs according to their final/enhanced scheduling found after using the segmented queue genetic solution.
\end{tablenotes}
\end{threeparttable}}
\end{center}
\end{table*}

In contrast, the scheduling approach with the segmented queue genetic solution has been applied on each individual queue of the tier to seek an optimal scheduling of jobs in that queue. The results (reported in Figure~\ref{fig:Case2_Queue_SYSTEM} and Table~2) demonstrate the effectiveness of this scheduling approach in optimizing the performance of the job schedule of $77$ jobs in the environment so as to reduce the service-level violation time and SLA penalty. Tier $T\!_1$ is allocated $34$ jobs distributed into $12$, $10$, and $12$ jobs in the resource queues $Q_{1,1}$, $Q_{1,2}$, and $Q_{1,3}$, respectively. On the other side, tier $T\!_2$ contains $43$ jobs whereby $Q_{2,1}$ is allocated $12$ jobs, $Q_{2,2}$ $10$ jobs, and $Q_{2,3}$ $12$ jobs.

The initial schedule of the $77$ jobs in resource queues of both tiers has at the beginning started with $333$ units of violation time at the multi-tier level of the environment, as shown in Figure~\ref{fig:SYSTEM_333_37}. Then, the scheduling approach with the segmented queue genetic setup has been applied on each individual queue of each tier. This scheduling approach has formed an enhanced scheduling of jobs in each queue that has reduced, at the multi-tier level, the total service-level violation time of jobs by $45\%$ to reach $181$ units of violation time. As a result, the total SLA violation penalty payable by the service provider has been optimized by $38.5\%$, a reduction from $2.537$ for the initial scheduling to $1.56$ for the enhanced scheduling of jobs.

Similar observations are in order with respect to improving the total service-level violation time and SLA penalty of each individual resource-queue in each tier as a result of employing the segmented queue genetic solution. For instance, the resource-queue $Q_{1,1}$ of tier $T\!_1$ shown in Figure~\ref{fig:Server1Tier1SYSTEM_38_9302} contains 10 jobs, but its total service-level violation time and penalty is reduced by $72\%$ and $65.6\%$, respectively.

Thus, the system virtualized queue and segmented queue genetic solutions have efficiently explored a big solution search space using a small number of genetic iterations to achieve such enhancements. Figure~\ref{fig:Tier1SYSTEM_84_5951} shows that the system virtualized queue required a total of only $1,\!000$ genetic iterations to efficiently seek an optimal schedule of jobs in tier $T_1$, each iteration employs $10$ chromosomes to evolve the optimal schedule. As such, $\nolinebreak{10\!\times\!10^{3}}$ scheduling orders are constructed and genetically manipulated throughout the search space, as opposed to $21!$ (approximately $\nolinebreak{5\!\times\!10^{19}}$) scheduling orders if a brute-force search strategy is employed to seek the optimal scheduling of jobs. Similar observations are in order with respect to the results reported on the segmented queue genetic solution.

\subsection{QoS Penalty Scheduling Evaluation of the Differentiated Waiting Time $\omega\!\mathcal{P\!T}\!\!_{i,j}$}

\begin{figure*}[!h]
\captionsetup{justification=centering}
        \centering
        \begin{subfigure}{0.32\textwidth}
		        \includegraphics[width=\textwidth]
                {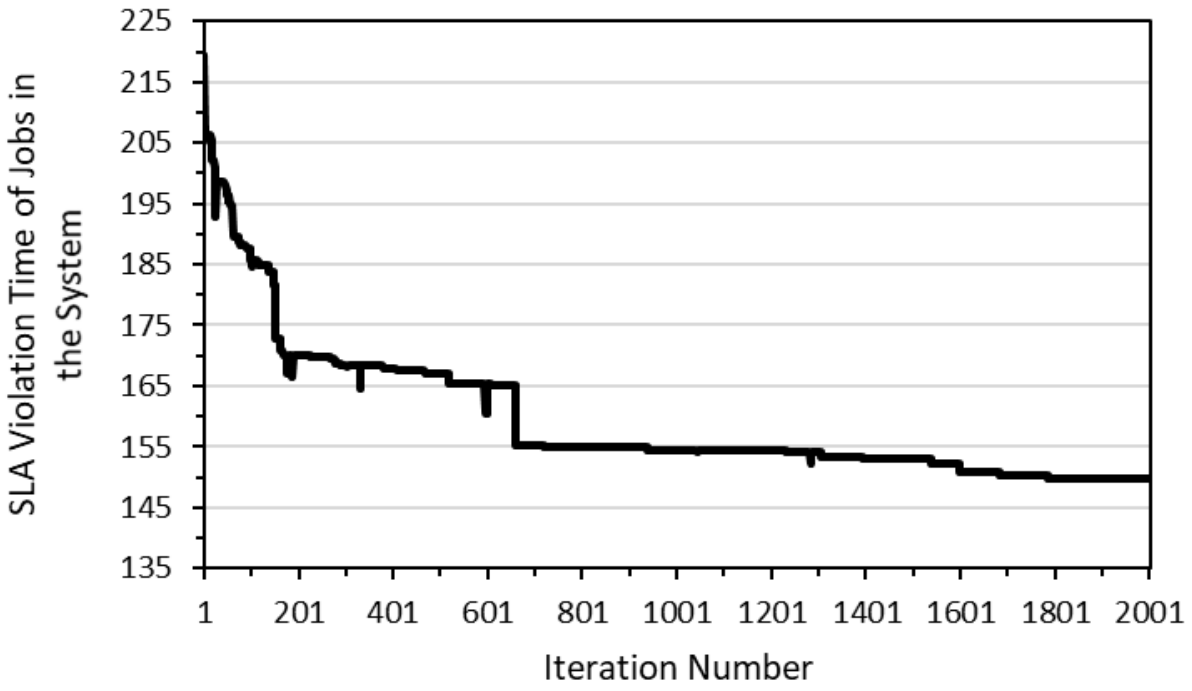}
	            \caption{System-Level (Total of 58 Jobs)}
	            \label{fig:SYSTEMproportional_219_5337}
        \end{subfigure}
        ~
        \begin{subfigure}{0.32\textwidth}
		        \includegraphics[width=\textwidth]
                {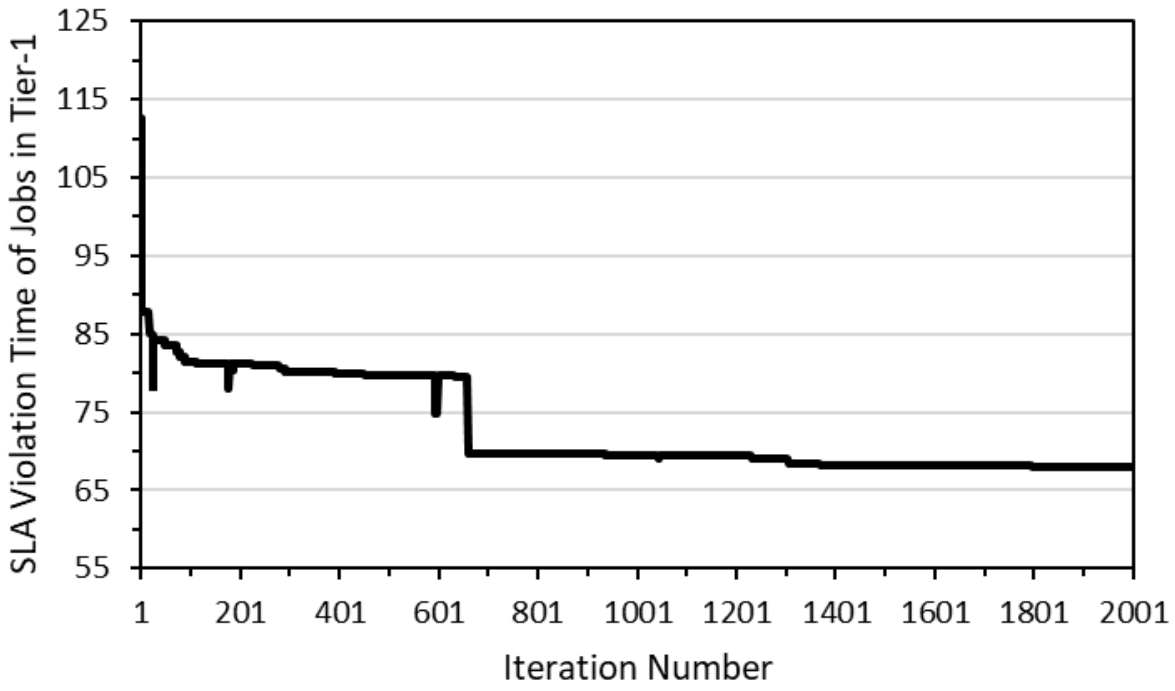}
	            \caption{Tier-1 (26 Jobs)}
	            \label{fig:Tier1SYSTEMproportional_112_4682}
        \end{subfigure}
        ~
        \begin{subfigure}{0.32\textwidth}
                \includegraphics[width=\textwidth]
                {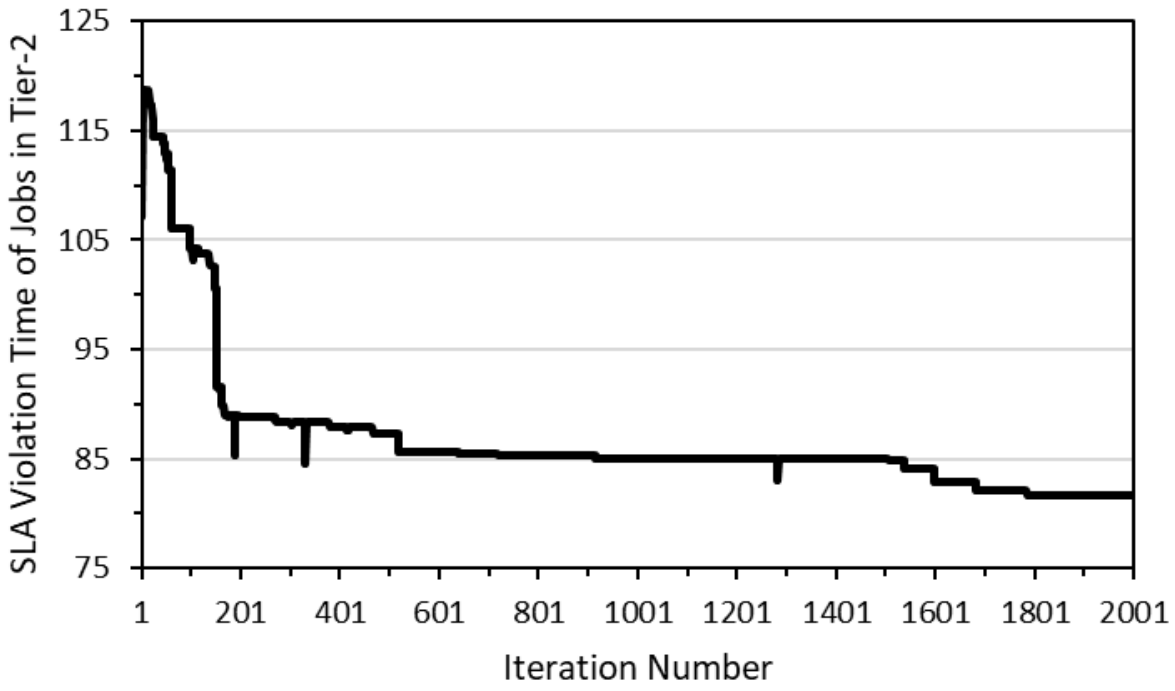}
	            \caption{Tier-2 (32 Jobs)}
	            \label{fig:Tier2SYSTEMproportional_107_0655}
        \end{subfigure}

        \caption{System Virtualized Queue Scheduling with Respect to Differentiated $\omega\!\mathcal{P\!T}\!_{i,j}$}
        \label{fig:Case2_Tier_SYSTEMproportional}
\end{figure*}

\begin{figure*}[!t]
\captionsetup{justification=centering}
        \centering
        \begin{subfigure}{0.32\textwidth}
		        \includegraphics[width=\textwidth]
                {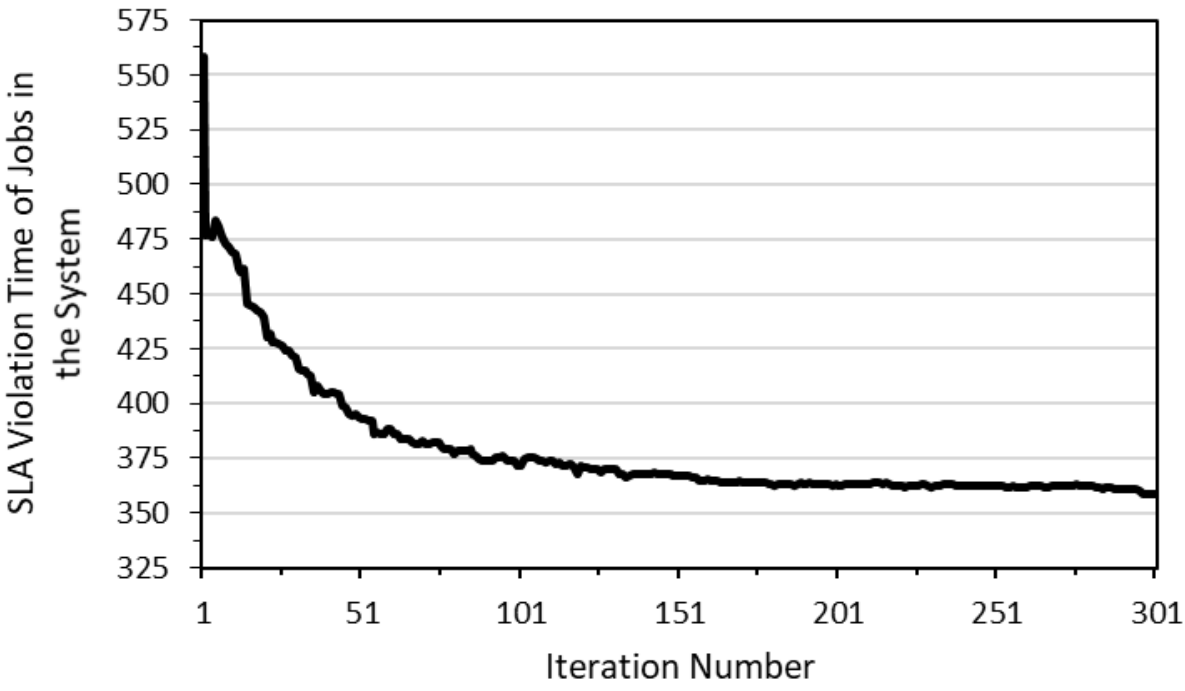}
	            \caption{System-Level (Total of 109 Jobs)}
	            \label{fig:SYSTEMproportional_558_327}
        \end{subfigure}

        \begin{subfigure}{0.32\textwidth}
		        \includegraphics[width=\textwidth]
                {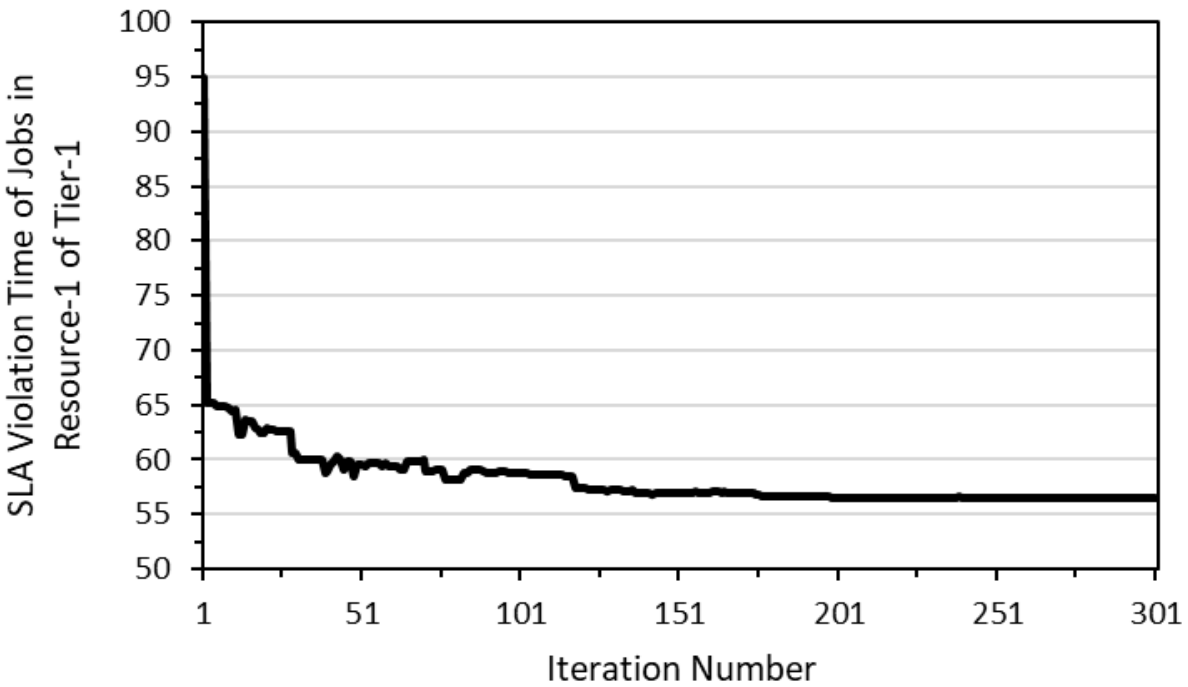}
	            \caption{Resource 1 of Tier 1 (Queue of 17 Jobs)}
	            \label{fig:Server1Tier1SYSTEMproportional_94_8766}
        \end{subfigure}
        ~
        \begin{subfigure}{0.32\textwidth}
		        \includegraphics[width=\textwidth]
                {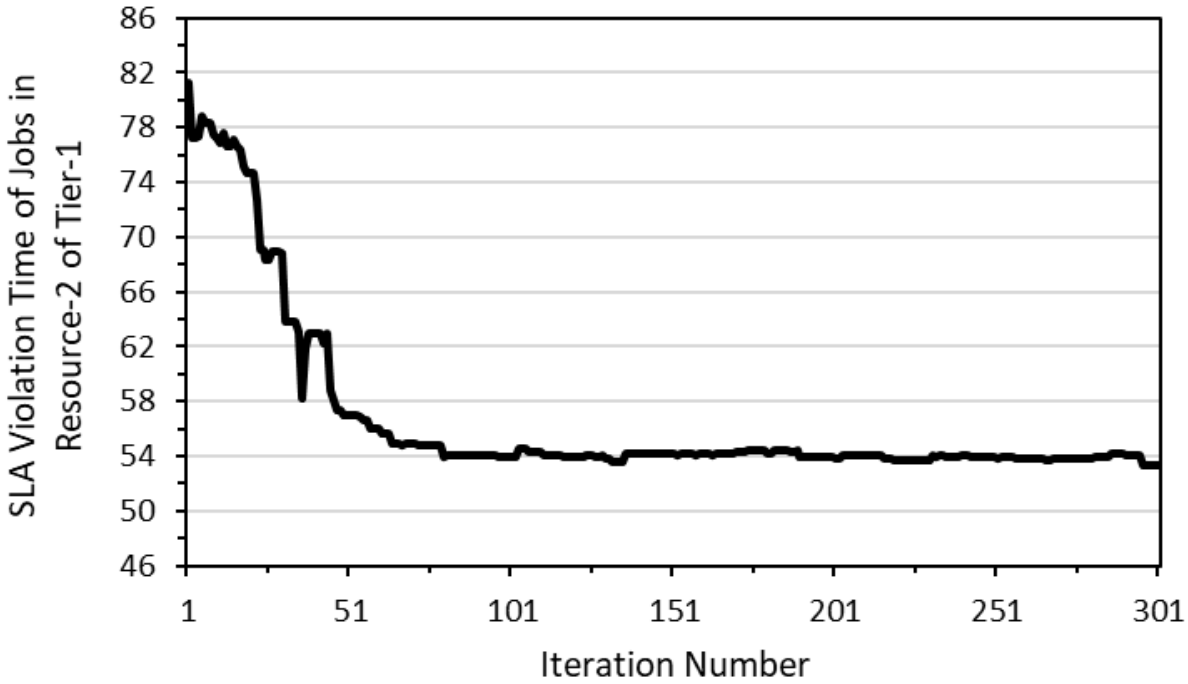}
	            \caption{Resource 2 of Tier 1 (Queue of 17 Jobs)}
	            \label{fig:Server2Tier1SYSTEMproportional_81_2750}
        \end{subfigure}
        ~
        \begin{subfigure}{0.32\textwidth}
                \includegraphics[width=\textwidth]
                {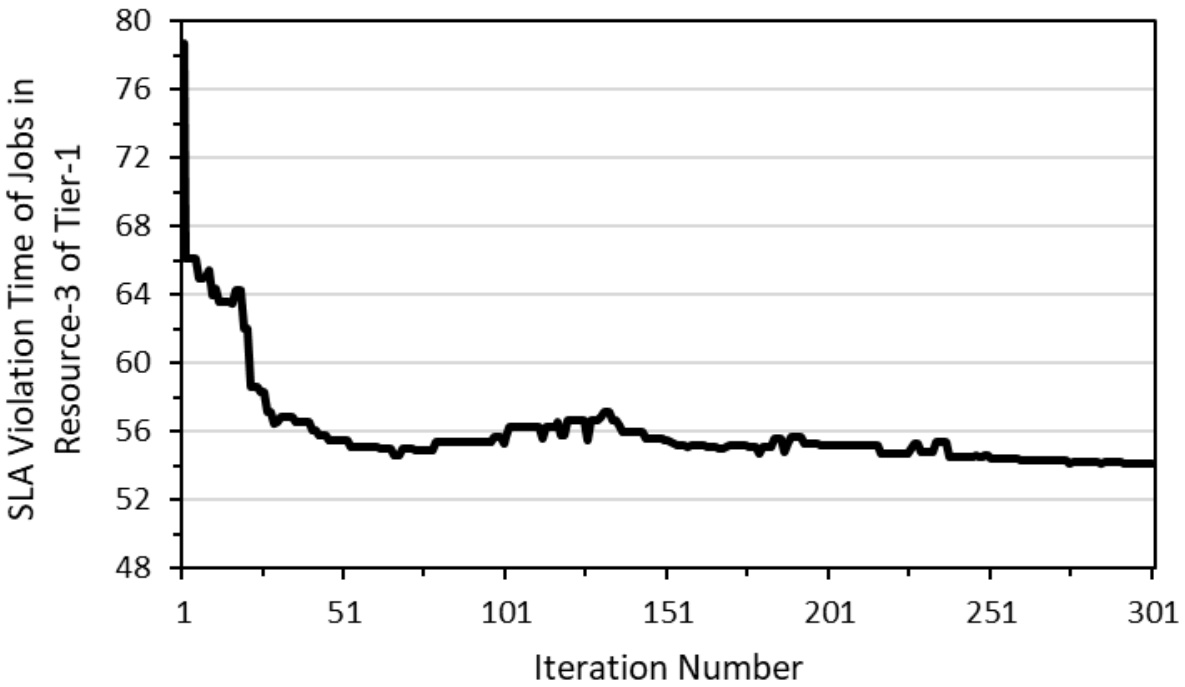}
	            \caption{Resource 3 of Tier 1 (Queue of 15 Jobs)}
	            \label{fig:Server3Tier1SYSTEMproportional_78_7112}
        \end{subfigure}

        \begin{subfigure}{0.32\textwidth}
		        \includegraphics[width=\textwidth]
                {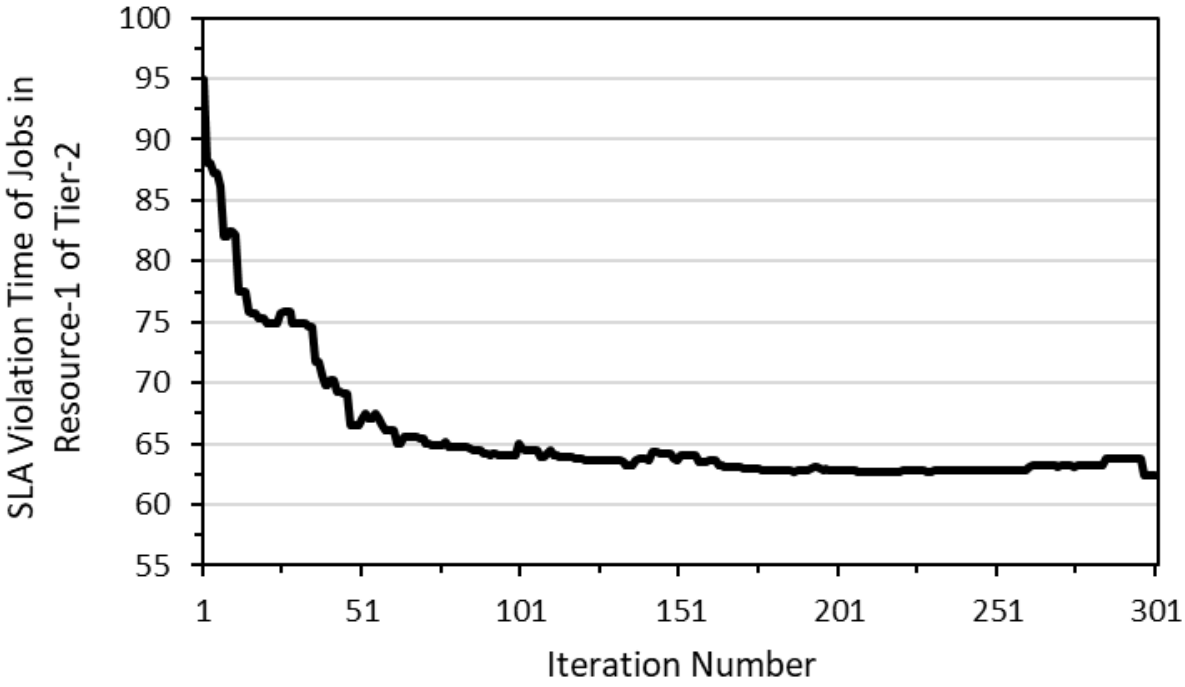}
	            \caption{Resource 1 of Tier 2 (Queue of 21 Jobs)}
	            \label{fig:Server1Tier2SYSTEMproportional_94_9211}
        \end{subfigure}
        ~
        \begin{subfigure}{0.32\textwidth}
		        \includegraphics[width=\textwidth]
                {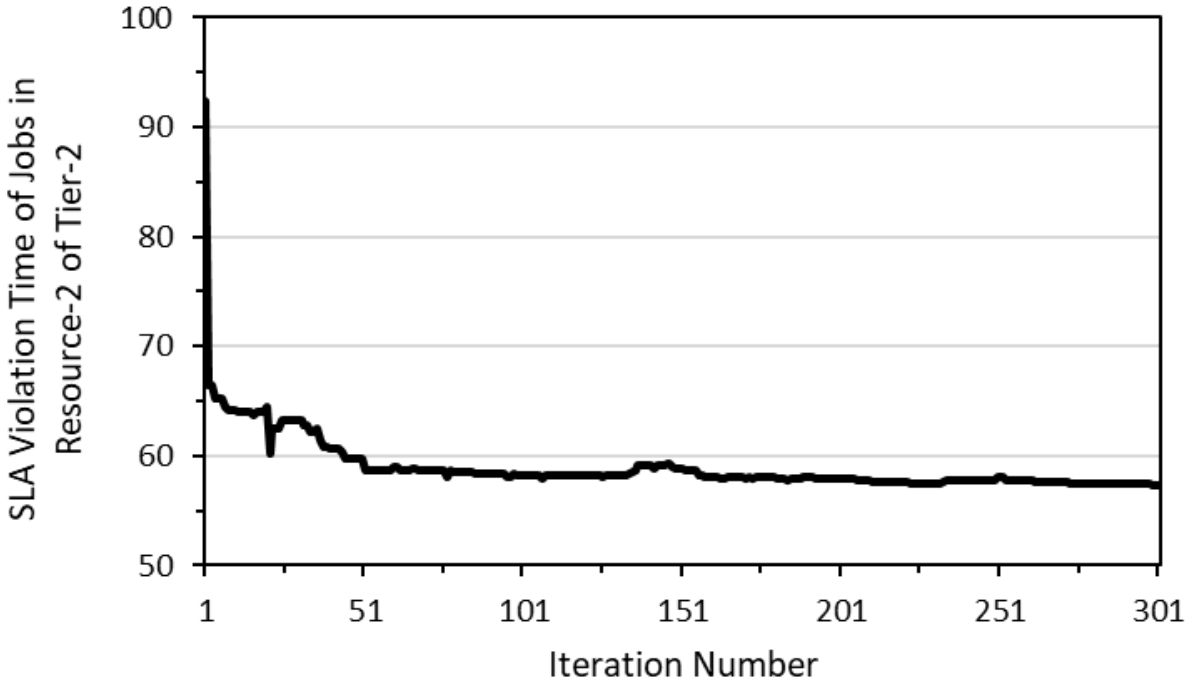}
	            \caption{Resource 2 of Tier 2 (Queue of 16 Jobs)}
	            \label{fig:Server2Tier2SYSTEMproportional_92_2903}
        \end{subfigure}
        ~
        \begin{subfigure}{0.32\textwidth}
                \includegraphics[width=\textwidth]
                {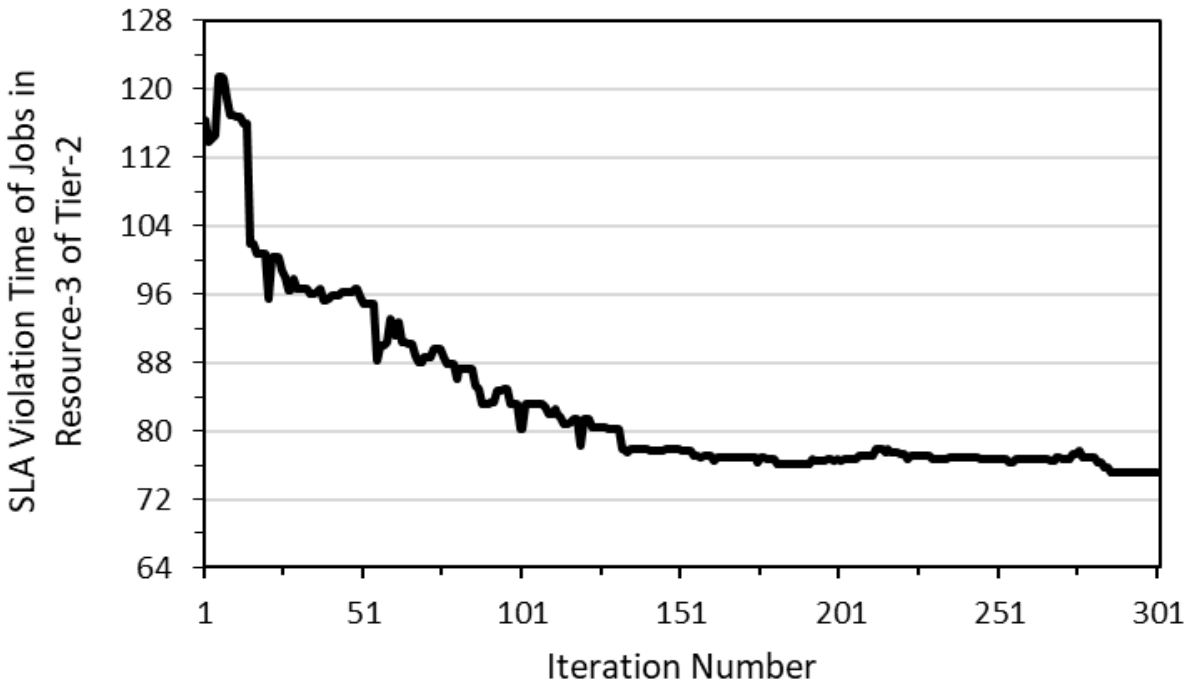}
	            \caption{Resource 3 of Tier 2 (Queue of 23 Jobs)}
	            \label{fig:Server3Tier2SYSTEMproportional_116_2528}
        \end{subfigure}

        \caption{Segmented Queue Scheduling with Respect to Differentiated $\omega\!\mathcal{P\!T}\!_{i,j}$}
        \label{fig:Case2_Queue_SYSTEMproportional}
\end{figure*}

The job schedules have been conducted according to the differentiated waiting time allowance $\omega\!\mathcal{P\!T}\!\!_{i,j}$ of each job $J_i$ at the tier level, which is derived from the waiting time allowance $\omega\!\mathcal{A\!L}_i$ of the job at the multi-tier level of the environment. Thus, the service-level violation time of each job $J_i$ is measured with respect to the $\omega\!\mathcal{P\!T}\!\!_{i,j}$ of the job in the tier, and accordingly the SLA violation penalty payable by the service provider is quantified. The system virtualized queue and segmented queue genetic solutions are used to efficiently seek optimal scheduling orders of jobs. Overall, the efficacy of the scheduling approach has been proven to produce optimal schedules that reduce the total service-level violation time of jobs and their associated SLA penalty at the multi-tier level of the environment (as shown in Figures~\ref{fig:Case2_Tier_SYSTEMproportional} and \ref{fig:Case2_Queue_SYSTEMproportional}, as well as Tables~3 and 4).

\begin{table*}[!b]
\label{tab:Tier12}
\captionsetup{justification=centering}
\caption{System Virtualized Queue Scheduling with Respect to Differentiated $\omega\!\mathcal{P\!T}\!_{i,j}$}
\begin{center}\scalebox{0.8}{
\begin{threeparttable}
\begin{tabular}{c|c|cccc|cc}
\hline
\multirow{2}{*}{}
& \multirow{2}{*}{\textbf{\begin{tabular}[c]{@{}c@{}}Number \\of Jobs\end{tabular}}}\tnote{1}
& \multicolumn{2}{c}{\textbf{Initial}\tnote{2}} & \multicolumn{2}{c|}{\textbf{Enhanced}\tnote{3}}
& \multicolumn{2}{c}{\textbf{Improvement}}  \\ \cline{3-8}
& & \textbf{Violation}  & \textbf{Penalty} & \textbf{Violation} & \textbf{Penalty} & \textbf{Violation \%} & \textbf{Penalty \%} \\ \hline \hline

System-Level, Figure~\ref{fig:SYSTEMproportional_219_5337}
& 58 & 219.53  & 1.34  & 149.62 & 1.05  & 31.85\%  & 21.64\%   \\ \hline
Tier-1, Figure~\ref{fig:Tier1SYSTEMproportional_112_4682}
& 26 & 112.47  & 0.68  & 68.03  & 0.49  & 39.51\%  & 26.91\%   \\
Tier-2, Figure~\ref{fig:Tier2SYSTEMproportional_107_0655}
& 32 & 107.07  & 0.66  & 81.58  & 0.56  & 23.80\%  & 15.14\%   \\ \hline

\end{tabular}
\begin{tablenotes}\scriptsize
\item[1] \textbf{Number of Jobs} represents the total number of jobs in queues of the tier/environment. For instance, the first entry (58 jobs) shows that the multi-tier environment contains 58 jobs in total. The second (21 jobs) and third (25 jobs) entries of the table mean that the 3 queues of tier-1 and tier-2 are allocated 26 and 32 jobs, respectively.
\item[2] \textbf{Initial Violation} represents the total SLA violation time of jobs according to their initial scheduling before using the system virtualized queue genetic solution.
\item[3] \textbf{Enhanced Violation} represents the total SLA violation time of jobs according to their final/enhanced scheduling found after using the system virtualized queue genetic solution.
\end{tablenotes}
\end{threeparttable}}
\end{center}
\end{table*}

Figure~\ref{fig:SYSTEMproportional_219_5337} and Table~3 represent a multi-tier environment that comprises $58$ jobs; $26$ jobs are allocated in tier $T\!_1$ and $32$ jobs are allocated in tier $T\!_2$. At the start, the schedule of the $58$ jobs in both tiers produced $219.5$ units of violation time. After the scheduling approach along with the system virtualized queue genetic solution is applied on the tiers, an enhanced schedule for the $58$ jobs in both tiers has been formed. Consequently, the service-level violation time of the enhanced scheduling of jobs is optimized at the multi-tier level by $31.85\%$ to reach $149.6$ units of violation time. As a result, the associated SLA violation penalty presented in Table~3 is optimized by $21.64\%$, a reduction from $1.34$ for the initial schedule to $1.05$ for the enhanced schedule of jobs. Similarly, such enhancements reduce the total violation time and SLA penalty of the virtual queue of each individual tier (as shown in Figures~\ref{fig:Tier1SYSTEMproportional_112_4682} and \ref{fig:Tier2SYSTEMproportional_107_0655}, as well as Table~3). For instance, the violation time and SLA penalty of the virtual-queue (26 jobs) of tier $T\!_1$ have respectively been reduced by $39.5\%$ and $26.9\%$, as shown in Figure~\ref{fig:Tier1SYSTEMproportional_112_4682}.

Furthermore, similar observations are in order with respect to the segmented queue genetic solution shown in Figure~\ref{fig:Case2_Queue_SYSTEMproportional} and Table~4, where the total service-level violation time and penalty of the $109$ jobs in the resource queues of both tiers are reduced at the multi-tier level by $35.7\%$ and $11\%$, respectively. Also, these enhancements affect the total violation time and penalty of the job schedules in each individual queue of each tier. For instance, the total violation time of $Q_{1,1}$ ($17$ jobs) shown in Figure~\ref{fig:Server1Tier1SYSTEMproportional_94_8766} is reduced by $40.5\%$, which accordingly reduced the SLA violation penalty of jobs in the queue by $29.5\%$.

\begin{table*}[!hb]
\label{tab:Queue1}
\captionsetup{justification=centering}
\caption{Segmented Queue Scheduling with Respect to Differentiated $\omega\!\mathcal{P\!T}\!_{i,j}$}
\begin{center}\scalebox{0.8}{
\begin{threeparttable}
\begin{tabular}{c|c|cccc|cc}
\hline
\multirow{2}{*}{} & \multirow{2}{*}{\textbf{\begin{tabular}[c]{@{}c@{}}Number \\of Jobs\end{tabular}}}
& \multicolumn{2}{c}{\textbf{Initial}\tnote{4}} & \multicolumn{2}{c|}{\textbf{Enhanced}\tnote{5}}
& \multicolumn{2}{c}{\textbf{Improvement}}  \\ \cline{3-8}
& & \textbf{Violation}  & \textbf{Penalty} & \textbf{Violation} & \textbf{Penalty} & \textbf{Violation \%} & \textbf{Penalty \%} \\ \hline \hline

System-Level, Figure~\ref{fig:SYSTEMproportional_558_327}
& 109 & 558.33 & 3.61 & 358.73 & 2.69 & 35.75\% & 25.49\%   \\ \hline

Resource-1 Tier-1, Figure~\ref{fig:Server1Tier1SYSTEMproportional_94_8766}
& 17 & 94.88 & 0.61  & 56.49 & 0.43 & 40.46\% & 29.57\%  \\
Resource-2 Tier-1, Figure~\ref{fig:Server2Tier1SYSTEMproportional_81_2750}
& 17 & 81.28 & 0.56  & 53.34 & 0.41 & 34.37\% & 25.70\%  \\
Resource-3 Tier-1, Figure~\ref{fig:Server3Tier1SYSTEMproportional_78_7112}
& 15 & 78.71 & 0.54  & 54.11 & 0.42 & 31.26\% & 23.30\%  \\ \hline

Resource-1 Tier-2, Figure~\ref{fig:Server1Tier2SYSTEMproportional_94_9211}
& 21 & 94.92 & 0.61  & 62.42  & 0.46 & 34.25\% & 24.25\%   \\
Resource-2 Tier-2, Figure~\ref{fig:Server2Tier2SYSTEMproportional_92_2903}
& 16 & 92.29 & 0.60  & 57.35  & 0.44 & 37.86\% & 27.58\%   \\
Resource-3 Tier-2, Figure~\ref{fig:Server3Tier2SYSTEMproportional_116_2528}
& 23 & 116.25 & 0.69 & 75.03  & 0.53 & 35.46\% & 23.21\%   \\ \hline

\end{tabular}
\begin{tablenotes}\scriptsize
\item[4] \textbf{Initial Violation} represents the total SLA violation time of jobs according to their initial scheduling before using the segmented queue genetic solution.
\item[5] \textbf{Enhanced Violation} represents the total SLA violation time of jobs according to their final/enhanced scheduling found after using the segmented queue genetic solution.
\end{tablenotes}
\end{threeparttable}}
\end{center}
\end{table*}

\subsection{Comparison of the Approaches}
\label{sec:compare}

\begin{figure}[!h]
\captionsetup{justification=centering}
\centering
      \includegraphics[width=0.7\textwidth]{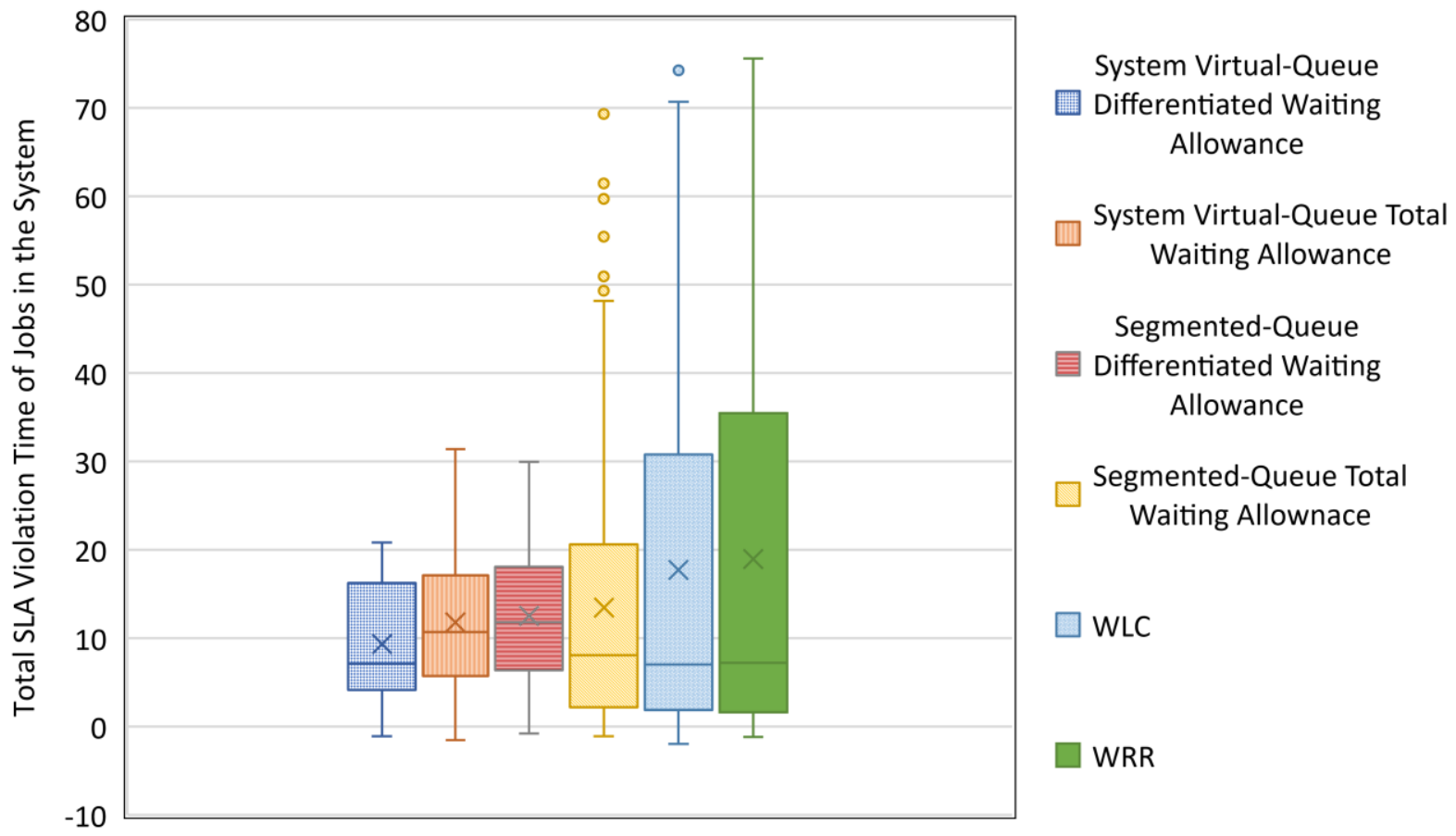}
	  \caption{Comparison of the Approaches}
      \label{fig:comparingThemTogether}
\end{figure}

Figure~\ref{fig:comparingThemTogether} and Table~5 contrast the performance of the scheduling approaches with respect to the total service-level violation time of jobs. The initial job schedules in the resource queues, and by implication, that of the system virtualized and segmented queues are the same. The WRR-based scheduling of jobs entails $3,\!812$ units of violation time, whilst the WLC-based scheduling entails $3,\!563$ units of violation time (as shown in Table~5). The scheduling approach along with the system virtualized queue and segmented queue genetic solutions has been applied to efficiently find optimized schedules that reduce the service-level violation time of jobs at the multi-tier level.

\begin{table*}[!ht]
\label{tab:SystemViolationPenalty}
\captionsetup{justification=centering}
\caption{Total SLA Violation Time}
\begin{center}\scalebox{0.7}{
\centering
\begin{tabular}{cc|cc|c|c}
\hline
 \multicolumn{2}{c|}{\textbf{\begin{tabular}[c]{@{}c@{}}Multi-Tier \\$\omega\!\mathcal{P\!T}\!\!_{i,j}$ Based Scheduling\end{tabular}}}
& \multicolumn{2}{c|}{\textbf{\begin{tabular}[c]{@{}c@{}}Multi-Tier \\$\omega\!\mathcal{A\!L}_i$ Based Scheduling\end{tabular}}}
& \multirow{3}{*}{\textbf{WLC}}
& \multirow{3}{*}{\textbf{WRR}} \\ \cline{1-4}
 \textbf{\begin{tabular}[c]{@{}c@{}}System \\ Virtualized Queue\end{tabular}}
& \textbf{Segmented Queue}
& \textbf{\begin{tabular}[c]{@{}c@{}}System \\ Virtualized Queue\end{tabular}}
& \textbf{Segmented Queue}  &  & \\ \hline

 1,859    & 2,495     & 2,363    & 2,700    & 3,563    & 3,812       \\ \hline

\end{tabular}}
\end{center}
\end{table*}

%
%
%
%

The multi-tier based scheduling with respect to the total waiting allowance $\omega\!\mathcal{A\!L}_i$ along with the segmented queue genetic solution entails $2,\!700$ units of violation time, a $29\%$ reduction compared with the WRR strategy and $24\%$ reduction compared with the WLC strategy. For the system virtualized queue genetic setup, the multi-tier $\omega\!\mathcal{A\!L}_i$ based scheduling produces job schedules that entail $2,\!363$ units of violation time, which is a reduction of $38\%$ compared with the WRR strategy and $34\%$ compared with the WLC strategy.

In contrast, the multi-tier based scheduling with respect to the differentiated waiting time allowance $\omega\!\mathcal{P\!T}\!\!_{i,j}$ generally produces better performance than the multi-tier $\omega\!\mathcal{A\!L}_i$ based scheduling. The $\omega\!\mathcal{P\!T}\!\!_{i,j}$ based scheduling along with the system virtualized queue genetic solution has produced job schedules that entail $1,\!859$ units of violation time, a reduction of $51\%$ compared with the WRR strategy and $48\%$ compared with the WLC strategy. On the other side of using the segmented queue genetic solution, the $\omega\!\mathcal{P\!T}\!\!_{i,j}$ based scheduling entails $2,\!495$ units of violation time, which gets $35\%$ and $30\%$ reductions compared with the WRR and WLC strategies, respectively.

Figure~\ref{fig:comparingThemTogether} depicts the average and maximum waiting performance of the scheduling strategies. Though, the $\omega\!\mathcal{P\!T}\!\!_{i,j}$ based scheduling along with the system virtualized queue genetic strategy shows the shortest average violation time and, therefore, the best performance among all the strategies; approximately an average of $9$ units of service-level violation time. Using the segmented queue genetic solution, the $\omega\!\mathcal{P\!T}\!\!_{i,j}$ based scheduling produces $13$ units of average service violation time, which is close to the multi-tier $\omega\!\mathcal{A\!L}_i$ based scheduling along with the system virtualized queue genetic solution that shows approximately $14$ units of average violation time. Nevertheless, the WRR and WLC job scheduling strategies delivered inferior performance.

Furthermore, similar observations are in order with respect to the maximum waiting performance. The WRR and WLC scheduling strategies produce the highest values of the maximum violation time of jobs, approximately $37$ units of violation time for the WRR and $32$ units of violation time for the WLC. The $\omega\!\mathcal{P\!T}\!\!_{i,j}$ based scheduling along with the system virtualized queue genetic strategy delivers the best performance in minimizing the total service-level violation time and thus the lowest SLA penalty; a maximum of $16$ units of violation time.

\section{Conclusion}
\label{sec:conc}

This paper presents a penalty-driven approach that addresses the optimal scheduling and allocation of jobs of various QoS obligations and computational demands in a multi-tier cloud environment. The approach employs the job's waiting time and service-level violation time to measure the penalty payable due to SLA violations, thus establishes a multi-tier-driven framework for quantifying and facilitating the management of a penalty that a cloud service provider can utilize to formulate penalty-based schedules. 

The scheduling approach contemplates the impact of schedules optimized in a given tier on the performance of schedules on subsequent tiers. The approach accounts for dependencies between tiers of the cloud environment to produce minimum penalty schedules at the multi-tier level. The performance of job schedules in a tier is optimized such that the potential of shifting and escalation of SLA violation penalties are mitigated when jobs progress through subsequent tiers.

The multi-tier-based biologically inspired genetic algorithm efficiently facilitates optimal scheduling of jobs, in a reasonable time. System virtualized and segmented queue abstractions mitigate the operator complexities of the scheduling process at the multi-tier level. Each queue abstraction represents a realization of an execution scheduling order of jobs. The virtualized abstraction collapses and reduces the solution search spaces of all queues of the multi-tier environment into a simple search space with one searching operator, that helps using the PGA efficiently seek optimal job schedules at the multi-tier level. 

The scheduling approach employs the multi-tier waiting time allowance $\omega\!\mathcal{A\!L}_i$ and the differentiated waiting time allowance $\omega\!\mathcal{P\!T}\!\!_{i,j}$ of each job to make multi-tier-driven scheduling decisions. Both experiments demonstrate the efficacy of the scheduling approach in optimizing the performance of job schedules, thus minimizing the service-level violation time and penalty payable by the cloud service provider at the multi-tier level. This scheduling approach with respect to both types of waiting time allowances, along with the system virtualized queue genetic solution, produces superior performance compared with the WRR and WLC scheduling strategies.

\section{Future Work}
\label{sec:fut}

The penalty model presented in this paper treats the violation penalty of different job waiting times to be identical. In fact, jobs of equal waiting times might not necessarily be similar in QoS penalty as such jobs tend to have different sensitivities to waiting and SLA violation. Therefore, it is imperative to design a penalty model that accounts for various QoS penalty classes, so that the performance of schedules is optimized at the tier and multi-tier levels to reflect such sensitivities.

\bibliographystyle{IEEEtran}

\bibliography{\jobname} 

\begin{thebibliography}{10}
\providecommand{\url}[1]{#1}
\csname url@samestyle\endcsname
\providecommand{\newblock}{\relax}
\providecommand{\bibinfo}[2]{#2}
\providecommand{\BIBentrySTDinterwordspacing}{\spaceskip=0pt\relax}
\providecommand{\BIBentryALTinterwordstretchfactor}{4}
\providecommand{\BIBentryALTinterwordspacing}{\spaceskip=\fontdimen2\font plus
\BIBentryALTinterwordstretchfactor\fontdimen3\font minus
  \fontdimen4\font\relax}
\providecommand{\BIBforeignlanguage}[2]{{%
\expandafter\ifx\csname l@#1\endcsname\relax
\typeout{** WARNING: IEEEtran.bst: No hyphenation pattern has been}%
\typeout{** loaded for the language `#1'. Using the pattern for}%
\typeout{** the default language instead.}%
\else
\language=\csname l@#1\endcsname
\fi
#2}}
\providecommand{\BIBdecl}{\relax}
\BIBdecl

\bibitem{NIST_CC}
R.~Bohn, J.~Messina, F.~Liu, J.~Tong, and J.~Mao, ``{NIST} cloud computing
  reference architecture,'' in \emph{Proceedings of the IEEE World Congress on
  Services}, July 2011, pp. 594--596.

\bibitem{An_analysis_LB2016}
C.~Thingom, G.~Kumar, and G.~Yeon, ``An analysis of load balancing algorithms
  in the cloud environment,'' in \emph{Proceedings of the International
  Conference on Communication and Electronics Systems}, October 2016, pp. 1--8.

\bibitem{cloudBigPic_2015}
D.~Puthal, B.~Sahoo, S.~Mishra, and S.~Swain, ``Cloud computing features,
  issues, and challenges: {A} big picture,'' in \emph{Proceedings of the
  International Conference on Computational Intelligence and Networks}, January
  2015, pp. 116--123.

\bibitem{Vinay2016}
V.~Chavan, K.~Dhole, and P.~Kaveri, ``Dynamic selection of job scheduling
  policies for performance improvement in cloud computing,'' in
  \emph{Proceedings of the International Conference on Computing for
  Sustainable Global Development}, March 2016, pp. 379--382.

\bibitem{nextGenerCl2018}
B.~Varghese and R.~Buyya, ``Next generation cloud computing: {New} trends and
  research directions,'' \emph{Future Generation Computer Systems}, vol.~79,
  pp. 849--861, 2018.

\bibitem{Mustafa2015}
S.~Mustafa, B.~Nazir, A.~Hayat, A.~Khan, and S.~Madani, ``Resource management
  in cloud computing: {Taxonomy}, prospects, and challenges,'' \emph{Computers
  \& Electrical Engineering}, vol.~47, no.~10, pp. 186--203, 2015.

\bibitem{AbdelzahirMap2015}
A.~Abdelmaboud, D.~Jawawi, I.~Ghani, A.~Elsafi, and B.~Kitchenham, ``Quality of
  service approaches in cloud computing: {A} systematic mapping study,''
  \emph{Journal of Systems and Software}, vol. 101, no.~3, pp. 159--179, 2015.

\bibitem{Nuaimi2012}
K.~Nuaimi, N.~Mohamed, M.~Nuaimi, and J.~Al-Jaroodi, ``A survey of load
  balancing in cloud computing: {Challenges} and algorithms,'' in
  \emph{Proceedings of the Symposium on Network Cloud Computing and
  Applications}, December 2012, pp. 137--142.

\bibitem{TaxonThakur2017}
A.~Thakur and M.~Goraya, ``A taxonomic survey on load balancing in cloud,''
  \emph{Journal of Network and Computer Applications}, vol.~98, no.~11, pp.
  43--57, 2017.

\bibitem{A_Survey_on_Scheduling_2014}
S.~Shaw and A.~Singh, ``A survey on scheduling and load balancing techniques in
  cloud computing environment,'' in \emph{Proceedings of the International
  Conference on Computer and Communication Technology}, September 2014, pp.
  87--95.

\bibitem{heurBalanc2015}
K.~Bey, F.~Benhammadi, and R.~Benaissa, ``Balancing heuristic for independent
  task scheduling in cloud computing,'' in \emph{Proceedings of the
  International Symposium on Programming and Systems}, April 2015, pp. 1--6.

\bibitem{SLAtree1}
Y.~Chi, H.~J. Moon, H.~Hacigumus, and J.~Tatemura, ``{SLA}-tree: {A} framework
  for efficiently supporting {SLA}-based decisions in cloud computing,'' in
  \emph{Proceedings of the International Conference on Extending Database
  Technology}, November 2011, pp. 129--140.

\bibitem{SoftHardSLA}
H.~Moon, Y.~Chi, and H.~Hacigumus, ``Performance evaluation of scheduling
  algorithms for database services with soft and hard {SLAs},'' in
  \emph{Proceedings of the Second International Workshop on Data Intensive
  Computing in the Clouds}, November 2011, pp. 81--90.

\bibitem{SaasWork2017}
G.~Stavrinides and H.~Karatza, ``The effect of workload computational demand
  variability on the performance of a {SaaS} cloud with a multi-tier {SLA},''
  in \emph{Proceedings of the IEEE International Conference on Future Internet
  of Things and Cloud}, August 2017, pp. 10--17.

\bibitem{GA_MinMin_2016}
S.~Rajput and V.~Kushwah, ``A genetic based improved load balanced {Min-Min}
  task scheduling algorithm for load balancing in cloud computing,'' in
  \emph{Proceedings of the International Conference on Computational
  Intelligence and Communication Networks}, December 2016, pp. 677--681.

\bibitem{MinMin_Priority_2013}
H.~Chen, F.~Wang, N.~Helian, and G.~Akanmu, ``User-priority guided {Min-Min}
  scheduling algorithm for load balancing in cloud computing,'' in
  \emph{Proceedings of the National Conference on Parallel Computing
  Technologies}, February 2013, pp. 1--8.

\bibitem{MinMin_2015}
G.~Patel, R.~Mehta, and U.~Bhoi, ``Enhanced load balanced {Min-Min} algorithm
  for static meta task scheduling in cloud computing,'' \emph{Procedia Computer
  Science}, vol.~57, no.~8, pp. 545--553, 2015.

\bibitem{MaxMin_2014}
X.~Li, Y.~Mao, X.~Xiao, and Y.~Zhuang, ``An improved {Max-Min} task-scheduling
  algorithm for elastic cloud,'' in \emph{Proceedings of the International
  Symposium on Computer, Consumer and Control}, June 2014, pp. 340--343.

\bibitem{Bianca_2004}
B.~Schroeder and M.~Harchol-Balter, ``Evaluation of task assignment policies
  for supercomputing servers: {The} case for load unbalancing and fairness,''
  \emph{Journal of Cluster Computing}, vol.~7, no.~2, pp. 151--161, 2004.

\bibitem{Mor_98}
M.~Harchol-Balter, M.~Crovella, and C.~Murta, ``On choosing a task assignment
  policy for a distributed server system,'' in \emph{Proceedings of the
  International Conference on Computer Performance Evaluation: {Modelling}
  Techniques and Tools}, September 1998, pp. 231--242.

\bibitem{unknownMaguluri2014}
S.~Maguluri and R.~Srikant, ``Scheduling jobs with unknown duration in
  clouds,'' \emph{IEEE/ACM Transaction on Networking}, vol.~22, no.~6, pp.
  1938--1951, 2014.

\bibitem{ShedRedundGARDNER2017}
K.~Gardner, M.~Harchol-Balter, E.~Hyytia, and R.~Righter, ``Scheduling for
  efficiency and fairness in systems with redundancy,'' \emph{Performance
  Evaluation}, vol. 116, no.~C, pp. 1--25, 2017.

\bibitem{Mor2017ReplicRedundant}
K.~Gardner, S.~Zbarsky, S.~Doroudi, M.~Harchol-Balter, E.~Hyytia, and
  A.~Scheller-Wolf, ``Queueing with redundant requests: {Exact} analysis,''
  \emph{Queueing Systems: Theory and Applications}, vol.~83, no. 3-4, pp.
  227--259, 2016.

\bibitem{replica_LB2016}
A.~Nahir, A.~Orda, and D.~Raz, ``Replication-based load balancing,'' \emph{IEEE
  Transactions on Parallel and Distributed Systems}, vol.~27, no.~2, pp.
  494--507, 2016.

\bibitem{Mor2016RedundantD1}
K.~Gardner, S.~Zbarsky, M.~Harchol-Balter, and A.~Scheller-Wolf, ``The power of
  {D} choices for redundancy,'' \emph{ACM Performance Evaluation Review},
  vol.~44, no.~1, pp. 409--410, 2016.

\bibitem{Mor2016RedundantD2}
K.~Gardner, S.~Zbarsky, M.~Velednitsky, M.~Harchol-Balter, and
  A.~Scheller-Wolf, ``Understanding response time in the redundancy-d system,''
  \emph{ACM Performance Evaluation Review}, vol.~44, no.~2, pp. 33--35, 2016.

\bibitem{WRR_2014}
W.~Wang and G.~Casale, ``Evaluating weighted round robin load balancing for
  cloud web services,'' in \emph{Proceedings of the International Symposium on
  Symbolic and Numeric Algorithms for Scientific Computing}, September 2014,
  pp. 393--400.

\bibitem{JSQ_delay_2017}
S.~Mehdian, Z.~Zhou, and N.~Bambos, ``Join-the-shortest-queue scheduling with
  delay,'' in \emph{Proceedings of the American Control Conference}, May 2017,
  pp. 1747--1752.

\bibitem{Randomized_JSQ_2016}
A.~Mukhopadhyay and R.~Mazumdar, ``Analysis of randomized
  join-the-shortest-queue ({JSQ}) schemes in large heterogeneous
  processor-sharing systems,'' \emph{IEEE Transactions on Control of Network
  Systems}, vol.~3, no.~2, pp. 116--126, 2016.

\bibitem{LBaaS2015}
P.-H. Liang and J.-M. Yang, ``Evaluation of two-level global load balancing
  framework in cloud environment,'' \emph{International Journal of Computer
  Science \& Information Technology}, vol.~7, no.~2, p.~1, 2015.

\bibitem{JIQ_2018}
C.~Wang, C.~Feng, and J.~Cheng, ``Distributed {Join-the-Idle-Queue} for low
  latency cloud services,'' \emph{IEEE/ACM Transactions on Networking},
  vol.~26, no.~5, pp. 2309--2319, 2018.

\bibitem{JIQ_multiDispatcher_2017}
M.~Boor, S.~Borst, and J.~Leeuwaarden, ``Load balancing in large-scale systems
  with multiple dispatchers,'' in \emph{Proceedings of the IEEE Conference on
  Computer Communications}, May 2017, pp. 1--9.

\bibitem{predict2010}
G.~Reig, J.~Alonso, and J.~Guitart, ``Prediction of job resource requirements
  for deadline schedulers to manage high-level {SLAs} on the cloud,'' in
  \emph{Proceedings of the IEEE International Symposium on Network Computing
  and Applications}, July 2010, pp. 162--167.

\bibitem{SAR2014}
P.~Hoang, S.~Majumdar, M.~Zaman, P.~Srivastava, and N.~Gael, ``Resource
  management techniques for handling uncertainties in user estimated job
  execution times,'' in \emph{Proceedings of the International Symposium on
  Performance Evaluation of Computer and Telecommunication Systems}, July 2014,
  pp. 626--633.

\bibitem{On_maximizing_2001}
Z.~Liu, M.~Squillante, and J.~Wolf, ``On maximizing service-level-agreement
  profits,'' in \emph{Proceedings of the ACM Conference on Electronic
  Commerce}, October 2001, pp. 213--223.

\bibitem{multi-dim2011}
H.~Goudarzi and M.~Pedram, ``Multi-dimensional {SLA}-based resource allocation
  for multi-tier cloud computing systems,'' in \emph{Proceedings of the IEEE
  International Conference on Cloud Computing}, July 2011, pp. 324--331.

\bibitem{rahhali2018hybrid}
H.~Rahhali and M.~Hanoune, ``Hybrid heuristic algorithm for load balancing in
  the cloud,'' \emph{International Jounral Computer Science and Network
  Security}, vol.~18, no.~4, pp. 109--115, 2018.

\bibitem{Maximizing_2011}
H.~Goudarzi and M.~Pedram, ``Maximizing profit in cloud computing system via
  resource allocation,'' in \emph{Proceedings of the International Conference
  on Distributed Computing Systems Workshops}, June 2011, pp. 1--6.

\bibitem{SLA_Based_Profit_2004}
L.~Zhang and D.~Ardagna, ``{SLA} based profit optimization in autonomic
  computing systems,'' in \emph{Proceedings of the International Conference on
  Service Oriented Computing}, November 2004, pp. 173--182.

\bibitem{multiObj2015}
L.~Zuo, L.~Shu, S.~Dong, C.~Zhu, and T.~Hara, ``A multi-objective optimization
  scheduling method based on the ant colony algorithm in cloud computing,''
  \emph{IEEE Access}, vol.~3, no.~12, pp. 2687--2699, 2015.

\bibitem{GA3}
Y.~Xiaomei, Z.~Jianchao, L.~Jiye, and L.~Jiahua, ``A genetic algorithm for job
  shop scheduling problem using co-evolution and competition mechanism,'' in
  \emph{Proceedings of the International Conference on Artificial Intelligence
  and Computational Intelligence}, October 2010, pp. 133--136.

\bibitem{GaTabu1}
X.~Li and L.~Gao, ``An effective hybrid genetic algorithm and tabu search for
  flexible job shop scheduling problem,'' \emph{International Journal of
  Production Economics}, vol. 174, no.~4, pp. 93--110, 2016.

\bibitem{heuristicPSO}
M.~Nouiri, A.~Bekrar, A.~Jemai, S.~Niar, and A.~Ammari, ``An effective and
  distributed particle swarm optimization algorithm for flexible job-shop
  scheduling problem,'' \emph{Journal of Intelligent Manufacturing}, vol.~29,
  no.~3, pp. 603--615, 2018.

\bibitem{PerfEval_2016}
T.~Atmaca, T.~Begin, A.~Brandwajn, and H.~Castel-Taleb, ``Performance
  evaluation of cloud computing centers with general arrivals and service,''
  \emph{IEEE Transactions on Parallel and Distributed Systems}, vol.~27, no.~8,
  pp. 2341--2348, 2016.

\end{thebibliography}
\end{document}